\documentclass[%
 reprint,
 superscriptaddress,
 amsmath,amssymb,
 aps,
 prapplied,
]{revtex4-2}

\usepackage{graphicx}% Include figure files
\usepackage{dcolumn}% Align table columns on decimal point
\usepackage{bm}% bold math
\usepackage{multirow}
\usepackage[utf8]{inputenc}
\usepackage[T1]{fontenc}
\usepackage{mathptmx}

\begin{document}

\preprint{APS/123-QED}

% \title{A Physics-Informed Neural Framework for High-Precision Magnetic Characterization of Test Masses for the Taiji Gravitational Wave Mission}

\title{High-Precision Ground Characterization of Test-Mass Magnetic Properties for the Taiji Gravitational Wave Mission via a Physics-Informed Neural Framework}

\author{Chang Liu}
\altaffiliation{These authors contributed equally to this work.}
\affiliation{National Space Science Center, Chinese Academy of Sciences, Beijing 100190, China}
\affiliation{Center for Gravitational Wave Experiment, National Microgravity Laboratory, Institute of Mechanics, Chinese Academy of Sciences, Beijing 100190, China}
\affiliation{University of Chinese Academy of Sciences, Beijing 100049, China}

\author{Qiong Deng}
\altaffiliation{These authors contributed equally to this work.}
\affiliation{Center for Gravitational Wave Experiment, National Microgravity Laboratory, Institute of Mechanics, Chinese Academy of Sciences, Beijing 100190, China}
\affiliation{Lanzhou Center of Theoretical Physics, Lanzhou University, Lanzhou 730000, China}

\author{Huadong Li}
\affiliation{Changchun Institute of Optics, Fine Mechanics and Physics, Chinese Academy of Sciences, Changchun 130033, China}

\author{Liwei Yang}
\affiliation{National Space Science Center, Chinese Academy of Sciences, Beijing 100190, China}
\affiliation{University of Chinese Academy of Sciences, Beijing 100049, China}

\author{Xiaodong Peng}
\affiliation{National Space Science Center, Chinese Academy of Sciences, Beijing 100190, China}

\author{Ziren Luo}
\affiliation{Center for Gravitational Wave Experiment, National Microgravity Laboratory, Institute of Mechanics, Chinese Academy of Sciences, Beijing 100190, China}
\affiliation{Key Laboratory of Gravitational Wave Precision Measurement of Zhejiang Province, Hangzhou Institute for Advanced Study, UCAS, Hangzhou 310024, China}
\affiliation{Taiji Laboratory for Gravitational Wave Universe (Beijing/Hangzhou), University of Chinese Academy of Sciences (UCAS), Beijing 100049, China}

\author{Yuzhu Zhang}
\affiliation{National Space Science Center, Chinese Academy of Sciences, Beijing 100190, China}

\author{Chen Gao}
\affiliation{National Space Science Center, Chinese Academy of Sciences, Beijing 100190, China}

\author{Xiaotong Wei}
\affiliation{Center for Gravitational Wave Experiment, National Microgravity Laboratory, Institute of Mechanics, Chinese Academy of Sciences, Beijing 100190, China}

\author{Minghui Du}
\affiliation{Center for Gravitational Wave Experiment, National Microgravity Laboratory, Institute of Mechanics, Chinese Academy of Sciences, Beijing 100190, China}

\author{Zihao Xiao}
\affiliation{Center for Gravitational Wave Experiment, National Microgravity Laboratory, Institute of Mechanics, Chinese Academy of Sciences, Beijing 100190, China}

\author{Peng Xu}
\email{xupeng@imech.ac.cn}
\affiliation{Center for Gravitational Wave Experiment, National Microgravity Laboratory, Institute of Mechanics, Chinese Academy of Sciences, Beijing 100190, China}
\affiliation{Key Laboratory of Gravitational Wave Precision Measurement of Zhejiang Province, Hangzhou Institute for Advanced Study, UCAS, Hangzhou 310024, China}
\affiliation{Taiji Laboratory for Gravitational Wave Universe (Beijing/Hangzhou), University of Chinese Academy of Sciences (UCAS), Beijing 100049, China}
\affiliation{Lanzhou Center of Theoretical Physics, Lanzhou University, Lanzhou 730000, China}

\author{Bo Liang}
\email{liangbo22@mails.ucas.ac.cn}
\affiliation{Center for Gravitational Wave Experiment, National Microgravity Laboratory, Institute of Mechanics, Chinese Academy of Sciences, Beijing 100190, China} 
\affiliation{University of Chinese Academy of Sciences, Beijing 100049, China}

\author{Zhi Wang}
\email{wangzhi@ciomp.ac.cn}
\affiliation{Changchun Institute of Optics, Fine Mechanics and Physics, Chinese Academy of Sciences, Changchun 130033, China}

\author{Li-e Qiang}
\email{qianglie@nssc.ac.cn}
\affiliation{National Space Science Center, Chinese Academy of Sciences, Beijing 100190, China}

\date{\today}% It is always \today, today,

\begin{abstract}
Taiji is a gravitational wave detection mission in space initiated by the Chinese Academy of Sciences, which will open the millihertz window through a heliocentric triangular constellation of three drag-free spacecraft. Its ultimate sensitivity is determined partly by the residual acceleration noise of the gravitational reference sensors (GRS), within which the coupling between the test-mass and the fluctuating environmental magnetic field constitutes one of the key stray-force contributions. Following the path established by the LISA and TianQin teams, high-precision ground characterization of remanent magnetic moment $\vec{m}_r$ and volume susceptibility $\chi$ of the test masses is a central step in the Taiji pre-launch test program. A persistent challenge for this characterization is the non-stationary, colored background noise inherent to torsion-pendulum facilities, which systematically biases classical Ordinary Least Squares (OLS) and Kalman filter (KF) estimators. We propose an AI-enhanced Differentiable Weighted Least Squares (AI-WLS) framework that fuses a dilated one-dimensional residual network, acting as a dynamic noise evaluator, with a fully differentiable analytical physical solver. This architecture preserves the exact linear mapping from the magnetic parameters to the torque response while autonomously identifying and suppressing contaminated data segments. Validated on real measured noise from the Changchun Institute of Optics, Fine Mechanics and Physics torsion-pendulum facility developed for Taiji, which achieves a torque sensitivity of order $10^{-13}\,\mathrm{N\cdot m\,Hz^{-1/2}}$, the AI-WLS framework bounds the maximum absolute estimation errors at $4.46\times 10^{-10}\,\mathrm{A\cdot m^2}$ for $\vec{m}_r$ and $7.8\times 10^{-8}$ for $\chi$, satisfying Taiji's ground-test requirements on all these parameters simultaneously, whereas OLS and KF both fail under the realistic non-stationary background. Beyond magnetic characterization, the framework provides a general methodology for ground-based verification of stray-force coupling models of ultra-precision GRSs, supporting the upcoming Taiji pre-launch noise-model test campaign and readily transferable to similar precision-measurement tasks for LISA-like space-borne gravitational wave missions.
%“太极”计划是由中国科学院发起的一项空间引力波探测任务，它将通过由三颗无拖曳航天器组成的日心三角形星座来开启毫赫兹频段的引力波探测窗口。其极限灵敏度部分取决于引力参考传感器（GRS）的残余加速度噪声，其中检验质量与波动的环境磁场之间的耦合构成了关键的杂散力来源之一。沿着 LISA 和天琴团队确立的路线，对检验质量的剩磁矩 $\vec{m}_r$ 和体积磁化率 $\chi$ 进行高精度地面表征，是太极发射前测试项目中的核心步骤。这种表征面临的一个长期挑战是扭秤装置中固有的非平稳有色背景噪声，这会使经典的普通最小二乘法（OLS）和卡尔曼滤波（KF）估计量产生系统性偏差。为此，我们提出了一种人工智能增强的可微加权最小二乘（AI-WLS）框架，该框架将作为动态噪声评估器的空洞一维残差网络与完全可微的解析物理求解器融合在一起。这种架构保留了从磁性参数到扭矩响应的精确线性映射，同时能够自主识别并抑制受污染的数据段。我们在长春光学精密机械与物理研究所为太极研制的扭秤设施（其扭矩灵敏度达到 $10^{-13}\,\mathrm{N\cdot m\,Hz^{-1/2}}$ 量级）的真实测量噪声上对该框架进行了验证。结果表明，AI-WLS 框架将 $\vec{m}_r$ 的最大绝对估计误差限制在 $4.46\times 10^{-10}\,\mathrm{A\cdot m^2}$，将 $\chi$ 的误差限制在 $7.8\times 10^{-8}$，从而同时满足了太极对所有这些参数的地面测试要求，而 OLS 和 KF 在现实的非平稳背景噪声下均宣告失效。除了磁性表征之外，该框架还为超高精度引力参考传感器（GRS）杂散力耦合模型的地面验证提供了一种通用方法，不仅能为即将到来的太极发射前噪声模型测试任务提供支持，且易于推广至类 LISA 的空间引力波探测任务中的类似精密测量工作。
\end{abstract}

\maketitle

\section{\label{sec:level1}Introduction}

Since the first direct detection of a binary black hole coalescence, GW150914~\cite{Abbott2016GR}, the LIGO-Virgo-KAGRA (LVK) network has catalogued hundreds of compact binary mergers~\cite{Abbott2021GR,Abbott2023}, opening the era of gravitational-wave (GW) astronomy and providing precision tests of general relativity in the strong-field regime. These ground-based interferometers are, however, restricted to the $10$--$10^{3}$~Hz band due to seismic and control-loop noise, leaving the scientifically rich millihertz window ($10^{-4}$ to $1$~Hz) accessible only from space. This window hosts massive black hole binary coalescences, extreme mass-ratio inspirals, Galactic double white dwarf binaries, and cosmological stochastic backgrounds~\cite{cpl_42_8_081101,LiangRapidPE,Liang2025TowardsEA,liang2026fluxmcrapidhighfidelityinference,Caprini2019_SGWB}, motivating an international effort to develop space-borne GW antennas, including the European-led LISA mission~\cite{AmaroSeoane2017,Armano2016} and the Chinese programs of Taiji and TianQin~\cite{Hu2017,Luo2016,Milyukov2020}. 
Initiated by the Chinese Academy of Sciences (CAS), the Taiji mission~\cite{Hu2017,Luo2020} is one of the Chinese contributions to this effort and will deploy a heliocentric equilateral-triangular constellation of three drag-free spacecraft with $3 \times 10^{6}$~km arm length, targeting the $0.1\,\mathrm{mHz}$-$1\,\mathrm{Hz}$ band.  
Incident GWs induce picometer level relative displacements between the free-falling TMs, which are read out by inter-satellite laser interferometry~\cite{Du_2023}.
%自 GW150914~\cite{Abbott2016GR} 以来，LIGO-Virgo-KAGRA (LVK) 网络已编目了数百个致密双星合并事件~\cite{Abbott2021GR,Abbott2023}，开启了引力波 (GW) 天文学时代，并提供了强场机制下广义相对论的精密检验。然而，由于地震和控制回路噪声，这些地基干涉仪被限制在 $10$--$10^{3}$~Hz 频段。预期可以通过空间探测到 $10^{-4}$ 到 $1$~Hz 之间的窗口，该窗口寄宿着大质量黑洞双星、极端质量比螺旋、银河系双白矮星和宇宙随机背景~\cite{cpl_42_8_081101,LiangRapidPE,Liang2025TowardsEA,liang2026fluxmcrapidhighfidelityinference}。因此，已经提出了一代空间引力波天线以开启这个毫赫兹 (mHz) 窗口，包括 LISA~\cite{AmaroSeoane2017,Armano2016} 以及中国太极和天琴计划~\cite{Hu2017,Luo2016,Milyukov2020}。太极由中国科学院发起，将发射三艘无拖曳航天器，形成一个臂长为 $3 \times 10^{6}$~km 的日心正三角形，目标是 0.1~mHz-1~Hz 频段~\cite{Luo2020}。入射的引力波会在自由落体检验质量 (TMs) 之间诱导皮米级的相对位移，这些位移通过星间激光干涉测量进行读取~\cite{Du_2023}。

At the core of these space-borne LISA-like antennas is the Gravitational Reference Sensor (GRS), which ensures that the TMs follow near-perfect geodesic trajectories. The ultimate sensitivity of the detector is governed by the residual acceleration noise of the GRS. For Taiji, the residual acceleration noise of the free-falling TM must be below $3\times 10^{-15}\,\mathrm{m\,s^{-2}\,Hz^{-1/2}}$ across the science band. Achieving such extreme inertial stability requires the mitigation of a long list of stray-force contributions, including Brownian gas damping, actuator and capacitive cross-couplings, electrostatic patch potentials, radiometer and radiation-pressure effects associated with thermal gradients, and residual charge accumulation from cosmic rays. Among these, the interaction between the TM's magnetic properties and the fluctuating environmental magnetic field is an important contribution, which induces stray force as~\cite{Armano_2025}
%这些类似LISA的空间天线的核心是引力参考传感器（GRS），它能确保测试质量（TM）遵循近乎完美的测地线轨迹。探测器的极限灵敏度由GRS的残余加速度噪声决定。对于太极计划，在整个科学频段内，自由下落的测试质量的残余加速度噪声必须低于$3\times 10^{-15}\,\mathrm{m\,s^{-2}\,Hz^{-1/2}}$。实现如此极端的惯性稳定性需要减轻一系列杂散力带来的影响，包括布朗气体阻尼、致动器和电容交叉耦合、静电斑块电位、与热梯度相关的辐射计和辐射压力效应，以及宇宙射线带来的残余电荷积累。在这些因素中，测试质量的磁特性与波动的环境磁场之间的相互作用是一个重要因素，它引起的杂散力如下~\cite{Armano_2025}\begin{equation}\vec{F} = \iiint_V \left{ (\vec{m}_r\cdot\nabla)\vec{B} + \frac{\chi}{\mu_0}\left[(\vec{B}\cdot\nabla)\vec{B}\right] \right} dx,dy,dz,\label{eq:magforce}\end{equation}其中$V$是测试质量体积，$\vec{m}_r$是它的残余磁化强度，$\chi$是它的体积磁化率，$\mu_0$是真空磁导率，$\vec{B}$是局部磁感应强度。这些影响的显著性已经在CHAMP等任务中得到了证明。在CHAMP任务中，磁力矩器的激活和环境磁场与STAR加速度计的耦合被确定为加速度计模式下不可忽视的系统误差，必须在Level-1B数据处理期间明确建模并减去~\cite{Peterseim2012,Flury2008}。
\begin{equation}
\vec{F} = \iiint_V \left\{ (\vec{m}_r\cdot\nabla)\vec{B} + \frac{\chi}{\mu_0}\left[(\vec{B}\cdot\nabla)\vec{B}\right] \right\} dx\,dy\,dz,
\label{eq:magforce}
\end{equation}
where $V$ is the TM volume, $\vec{m}_r$ its remanent magnetization, $\chi$ its volume magnetic susceptibility, $\mu_0$ the vacuum permeability, and $\vec{B}$ the local magnetic induction. The significance of these effects has been demonstrated in missions such as the CHAMP mission, where the magnetic-torquer activation and ambient-field coupling to the STAR accelerometer were identified as non-negligible systematics in accelerometer mode, which had to be explicitly modeled and subtracted during Level-1B data processing~ \cite{Peterseim2012,Flury2008}. 
Consequently, the rigorous verification of magnetic coupling models and the high-precision characterization of $\vec{m}_r$ and $\chi$ are paramount.
Such measurements are not only critical for the hardware R\&D of the GRS to ensure material compliance but are also indispensable for post-processing data analysis, where accurate physical models allow for the subtraction of magnetic noise from the science signal.
%因此，严格验证磁耦合模型和高精度表征$\vec{m}_r$和$\chi$是至关重要的。这些测量不仅对GRS硬件研发至关重要（以确保材料合规性），而且对后处理数据分析也是必不可少的，其中精确的物理模型允许从科学信号中减去磁噪声。
%受这些要求驱动，LISA和天琴团队都使用高精度扭摆对测试质量磁特性进行了系统的地面表征。LISA要求$|\vec{m}_r| < 2\times 10^{-8},\mathrm{A\cdot m^2}$和$\chi < 3\times 10^{-6}$，而天琴采用了更严格的约束条件$|\vec{m}_r| < 10^{-8},\mathrm{A\cdot m^2}$和$\chi < 10^{-6}$~\cite。LISA探路者的飞行中测量已经证明了在这个磁清洁水平上制造测试质量的实际可行性，报告称$|\vec{m}_r| < 2.5\times 10^{-10},\mathrm{A\cdot m^2}$和$\chi < 3.37\times 10^{-5}$~\cite{Armano_2025}。

Motivated by these requirements, both the LISA and TianQin teams have pursued systematic ground-based characterization of TM magnetic properties using high-precision torsion pendulums~\cite{PhysRevApplied.15.014008,PhysRevLett.120.061101,Hueller_2005,PhysRevLett.91.151101,Antonucci_2011}. The LISA specifications~\cite{amaroseoane2017laserinterferometerspaceantenna} require $|\vec{m}_r| < 2\times 10^{-8}\,\mathrm{A\cdot m^2}$ and $\chi < 3\times 10^{-6}$, while TianQin adopts even tighter constraints $|\vec{m}_r| < 10^{-8}\,\mathrm{A\cdot m^2}$ and $\chi < 10^{-6}$~\cite{Luo2016,su2020}. In-flight measurements by LISA Pathfinder have demonstrated the practical feasibility of fabricating TMs at this level of magnetic cleanliness, reporting $|\vec{m}_r| < 2.5\times 10^{-10}\,\mathrm{A\cdot m^2}$ and $\chi < 3.4\times 10^{-5}$~\cite{Armano_2025}. Given Taiji's comparable magnetic budget, the pre-launch verification of its TMs targets resolutions of $1 \times 10^{-9}\,\mathrm{A\cdot m^2}$ for $|\vec{m}_r|$ and $1 \times 10^{-7}$ for $\chi$, ensuring the ground-test precision exceeds mission specifications by at least one order of magnitude to effectively mitigate the risk of measurement errors.
%受这些要求驱动，LISA和天琴团队都使用高精度扭摆对测试质量磁特性进行了系统的地面表征~\cite{PhysRevApplied.15.014008,PhysRevLett.120.061101,Hueller_2005,PhysRevLett.91.151101,Antonucci_2011}。LISA要求$|\vec{m}_r| < 2\times 10^{-8},\mathrm{A\cdot m^2}$和$\chi < 3\times 10^{-6}$，而天琴采用了更严格的约束条件$|\vec{m}_r| < 10^{-8},\mathrm{A\cdot m^2}$和$\chi < 10^{-6}$~\cite{Luo2016,su2020}。LISA探路者的飞行中测量已经证明了在这个磁清洁水平上制造测试质量的实际可行性，报告称$|\vec{m}_r| < 2.5\times 10^{-10},\mathrm{A\cdot m^2}$和$\chi < 3.37\times 10^{-5}$~\cite{Armano_2025}。鉴于太极类似的磁预算，其测试质量的发射前验证目标分辨率定为：$|\vec{m}_r|$ 为 1 x 10^-9 A·m^2，\chi 为 1 x 10^-7，以确保地面测试精度超过任务规范至少一个数量级，从而有效降低测量误差的风险。

% Because Taiji adopts a comparable magnetic budget, and because ground-test precision should exceed the mission specification by at least one order of magnitude, the pre-launch verification of Taiji's TMs targets resolutions of $1 \times 10^{-9}\,\mathrm{A\cdot m^2}$ for $|\vec{m}_r|$ and $1 \times 10^{-7}$ for $\chi$. 
% 因为太极采用了相当的磁预算，并且因为地面测试精度应该超过任务规范至少一个数量级，所以太极的测试质量的发射前验证目标分辨率对于$|\vec{m}_r|$为$1 \times 10^{-9},\mathrm{A\cdot m^2}$，对于$\chi$为$1 \times 10^{-7}$。
To support the Taiji GRS R\&D at this level, the Changchun Institute of Optics, Fine Mechanics and Physics (CIOMP) of the CAS has developed a dedicated weak-force torsion-pendulum facility with a torque sensitivity of order $10^{-13}\,\mathrm{N\cdot m\,Hz^{-1/2}}$. Jointly with the Institute of Mechanics and the National Space Science Center of CAS, the CIOMP team is now positioned to carry out the systematic verification of GRS noise models for Taiji.
%为了支持太极在此级别的GRS研发，中国科学院长春光学精密机械与物理研究所（CIOMP）开发了专用的弱力扭摆设施，其扭矩灵敏度约为$10^{-13},\mathrm{N\cdot m,Hz^{-1/2}}$。CIOMP团队现在与力学研究所和中国科学院国家空间科学中心合作，准备对太极的GRS噪声模型进行系统的验证。

At this stage, the CIOMP team has completed the commissioning of the torsion-pendulum facility and collected long-duration free-run data. These records exhibit the expected colored, non-stationary character of the experimental background, dominated by low-frequency drifts and transient glitches, which violates the stationary-Gaussian noise assumption underpinning both the OLS~\cite{LIU2023107048} and KF~\cite{liu2025design} estimators and can introduce systematic biases, especially given the strong non-linear cross-coupling between $\vec{m}_r$ and $\chi$. 
To push the measurement precision of the TM magnetic parameters toward the level required by Taiji under such realistic conditions, we propose in this work an end-to-end AI-enhanced Differentiable Weighted Least Squares (AI-WLS) framework that couples a deep dilated residual network with a fully differentiable analytical WLS solver. The network acts as a dynamic, data-driven noise evaluator, assigning time-varying confidence weights to the measured torque sequence so that contaminated segments are automatically down-weighted, while the WLS module preserves the exact linear physical mapping from the magnetic coupling parameters to the torque response; and because the full pipeline is differentiable, the parameter-estimation error backpropagates into the weight network. 
Given the demonstrated noise performance of the CIOMP pendulum, we find that this framework is able to reach Taiji's magnetic-characterization requirements across the range of expected experimental conditions examined in this work. Because a calibrated magnetic-torque injection with known $(\vec{m}_r, \chi)$ ground truth is not available at this commissioning stage, we validate AI-WLS by superposing physically consistent, numerically simulated magnetic-coupling signals onto the real ``free-run'' noise recorded by the pendulum, benchmarking it against the OLS and KF baselines under identical noise conditions. More broadly, the framework is designed as a general-purpose tool for ground-based verification of stray-force coupling models of the Taiji GRS, and is potentially applicable to other noise couplings beyond magnetism, such as thermal, radiometer, radiation-pressure, and charge-related torques, where the same torsion-pendulum methodology is employed. This paper thus provides the methodological foundation for the upcoming magnetic-coupling measurement campaign on the Taiji TMs, and contributes toward the Taiji GRS pre-launch noise-model verification program.
%在这个阶段，CIOMP团队已经完成了扭摆设施的调试，并收集了长期的自由运行数据。这些记录显示出实验背景预期的有色、非平稳特征，主要由低频漂移和瞬态干扰主导，这违反了普通最小二乘法（OLS）~\cite{LIU2023107048}和卡尔曼滤波~\cite{liu2025design}估计器所依赖的平稳高斯噪声假设，并引入了系统偏差，特别是考虑到$\vec{m}_r$和$\chi$之间强烈的非线性交叉耦合。为了在这样真实的条件下将TM磁参数的测量精度提高到太极所需的水平，我们在本研究中提出了一种端到端的AI增强可微加权最小二乘（AI-WLS）框架，该框架将深度扩张残差网络与完全可微的分析WLS求解器相结合。网络作为一个动态的、数据驱动的噪声评估器，将随时间变化的置信权重分配给测量的扭矩序列，以便受污染的段被自动降低权重，而WLS模块保留了从磁耦合参数到扭矩响应的精确线性物理映射，并且因为整个管道是可微的，参数估计误差可以端到端地反向传播到权重网络中。鉴于CIOMP扭摆已证明的噪声性能，我们认为该框架能够达到太极的磁特性要求。由于在这个调试阶段，具有已知$(\vec{m}_r, \chi)$真实值的校准磁扭矩注入尚未准备好，我们通过将物理上一致的、数值模拟的磁耦合信号叠加到扭摆记录的真实“自由运行”噪声上，来展示和验证AI-WLS框架，将其与在相同噪声条件下的OLS和卡尔曼滤波基线进行基准测试，从而为即将进行的太极TM测量活动奠定方法学基础。

\section{Experimental Setup}
\label{sec:experiment}

\subsection{The CIOMP Torsion Pendulum Facility}
\label{sec:pendulum}
\begin{figure}
\includegraphics[width=0.4\textwidth]{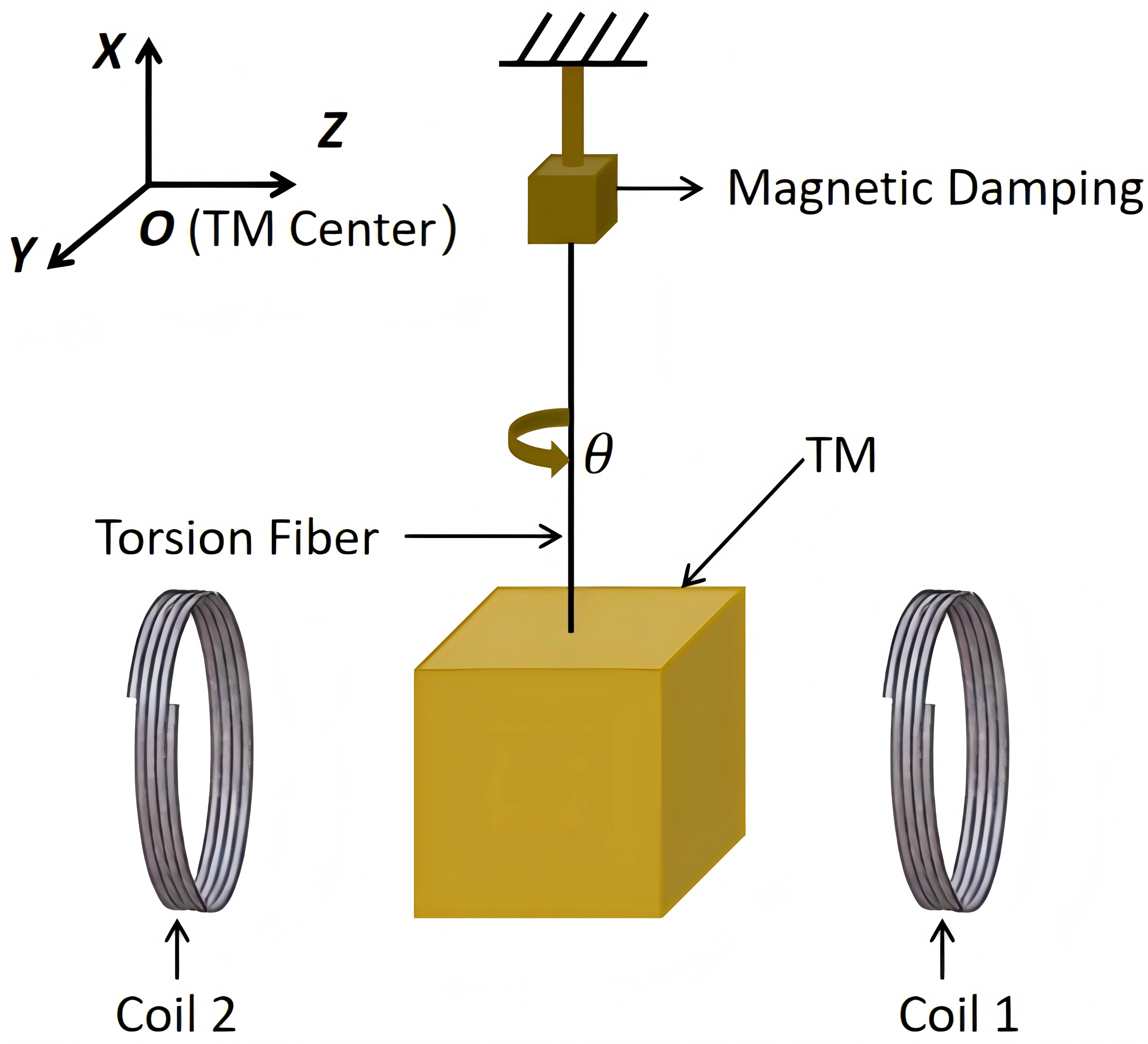}
\caption{\label{fig:pendulum} Schematic of the single-TM torsion pendulum. Two circular coils flanking the TM generate the driving magnetic field.}
%单检验质量（TM）扭摆示意图。位于检验质量两侧的两个圆形线圈负责产生驱动磁场。
\end{figure}
Fiber-suspended torsion pendulums are the established ground-test platforms for LISA-class GRS qualification, offering unparalleled sensitivity to ultra-weak torques in the sub-millihertz to hertz band~\cite{Russano2018,Hueller_2005,PhysRevApplied.15.014008,PhysRevLett.91.151101}. 
To support the magnetic characterization of the Taiji TM, a dedicated facility has been commissioned at CIOMP (Fig.~\ref{fig:pendulum}). The coordinate system is defined with the origin at the TM center, The $x$-axis is aligned parallel to the fiber, the $z$-axis is perpendicular to the coil plane, and the $y$-axis completes the right-handed Cartesian coordinate system. 
The setup features a solid copper-tungsten alloy cubic TM suspended by a fused-silica fiber along the vertical $x$-axis, with upper-stage magnetic damping to suppress residual swing modes. 
An autocollimator monitors the TM's rotation, making the instrument selectively responsive to $x$-axis torques. 
%纤维悬挂扭摆是 LISA 类引力参考传感器（GRS）鉴定的公认地面测试平台，在亚毫赫兹至赫兹频段内对超微弱转矩具有无与伦比的灵敏度。为了支持“太极”计划检验质量（TM）的磁特性表征，长春光机所（CIOMP）建成并启用了一套专用测试设施（图~\ref{fig:pendulum}）。坐标系的原点定义在 TM 中心，$x$ 轴与悬挂纤维平行，$z$ 轴垂直于线圈平面，$y$ 轴构成右手笛卡尔坐标系。该装置利用一根熔石英纤维沿垂直的 $x$ 轴悬挂一个实心铜钨合金立方体 TM，并配有上级磁阻尼结构以抑制残余的摆动模态。自准直仪用于监测 TM 的旋转，使该仪器能够选择性地响应沿 $x$ 轴方向的转矩。
Key geometric and physical parameters of the system are summarized in Table~\ref{tab:tm_parameters}. 
Crucially, the system's intrinsic torsional resonance frequency is engineered to be well below the magnetic-driving frequency. 
This ensures that the magnetic-coupling response is evaluated strictly within the mass-dominated regime, where the TM's rotational dynamics are governed by the standard linear damped-oscillator equation of motion.
%系统的主要几何与物理参数总结于表~\ref{tab:tm_parameters} 中。至关重要的是，系统的固有扭转共振频率被设计为远低于磁驱动频率。这确保了磁耦合响应能够严格在质量主导区间（mass-dominated regime）内进行评估，在此状态下，TM 的旋转动力学由标准的线性阻尼振子运动方程支配。

% --- 表格部分 ---
\begin{table}[htbp]
\centering
\caption{Physical and geometric parameters of the TM and its suspension system.}
\label{tab:tm_parameters}
\begin{tabular}{llc}
\hline\hline
Parameter & Symbol & Value \\
\hline
\multicolumn{3}{c}{\textit{Test Mass (Solid copper-tungsten alloy cube)}} \\
Side length & $L$ & $46\,\mathrm{mm}$ \\
Mass of TM & $M$ & $903\,\mathrm{g}$ \\
Moment of inertia ($x$-axis) & $I_x$ & $3.1474 \times 10^{-4}\,\mathrm{kg\cdot m^2}$ \\
\vspace{1mm} \\
\multicolumn{3}{c}{\textit{Suspension System (Quartz fiber)}} \\
Fiber length & $l$ & $\simeq 1\,\mathrm{m}$ \\
Fiber diameter & $d$ & $100\,\mathrm{\mu m}$ \\
Mechanical quality factor & $Q$ & $\simeq 10^5$ \\
Torsional resonance frequency & $f_0$ ($\omega_0/2\pi$) & $\simeq 0.0185\,\mathrm{Hz}$ \\
\hline\hline
\end{tabular}
\end{table}

\begin{equation}
I\,\ddot{\theta}(t) + \gamma\,\dot{\theta}(t) + \Gamma\,\theta(t) = N(t),
\label{eq:dynamics}
\end{equation}
where $\Gamma= I\omega_{0}^2$ is the torsional stiffness of the fiber, $N(t)$ is the external torque on the sensitive axis, and $\gamma = I\omega_{0}/Q$ is the internal damping coefficient. The angular displacement $\theta(t)$ is read out by the autocollimator at a sampling rate of $10\,\mathrm{Hz}$, and the torque time series $N(t)$ is reconstructed by applying the forward operator of Eq.~\eqref{eq:dynamics} to $\theta(t)$~\cite{Russano2018}.
%其中$k$为悬丝的扭转刚度，$N(t)$为作用在敏感轴上的外部力矩，$\gamma = I\omega_{0}/Q$为内部阻尼系数。角位移$\theta(t)$由自准直仪以 10 Hz 的采样率进行读取，并且通过将公式~\eqref{eq:dynamics}的正向算子应用于$\theta(t)$来重建力矩时间序列$N(t)$~\cite{Russano2018}。

At the working-point performance, the single-sided torque amplitude spectral density (ASD) of the pendulum background is well described as the incoherent sum of three dominant contributions~\cite{Hueller_2005,Russano2018},
\begin{equation}
S_{N}(\omega) = S_{N,\mathrm{th}}(\omega) + S_{N,\mathrm{ro}}(\omega) + S_{N,\mathrm{env}}(\omega),
\label{eq:noise_budget}
\end{equation}
where $S_{N,\mathrm{th}}$ is the thermal-noise floor set by the fluctuation--dissipation theorem applied to the internal friction of the fiber (dominant at low frequency and scaling as $\sqrt{4k_{B}T\,I\omega_{0}/Q}$ in the structural-damping limit), $S_{N,\mathrm{ro}}$ is the autocollimator readout contribution (approximately flat in angle, and therefore rising as $\omega^{2}$ when referred to torque, dominating the high-frequency band), and $S_{N,\mathrm{env}}$ collects the non-stationary environmental contributions such as seismic tilt, stray magnetic-field fluctuations, and thermal drifts of the surrounding structure; the last is known to exhibit diurnal and weekly modulations correlated with human activity near the test site. Commissioning runs with the driving coils un-energized yield a torque noise floor of order $10^{-13}\,\mathrm{N\cdot m\,Hz^{-1/2}}$ across the measurement band, meeting the torque-resolution requirement for magnetic-coupling verification of the Taiji TM at the target precision stated in Sec.~\ref{sec:level1}.
%在工作点性能下，扭摆背景的单边力矩振幅谱密度（ASD）可以很好地描述为三个主要贡献的不相干和~\cite{Hueller_2005,Russano2018}，\begin{equation}S_{N}(\omega) = S_{N,\mathrm{th}}(\omega) + S_{N,\mathrm{ro}}(\omega) + S_{N,\mathrm{env}}(\omega),\label{eq:noise_budget}\end{equation}其中 $S_{N,\mathrm{th}}$ 是由应用于悬丝内部摩擦的涨落耗散定理设定的热噪声基底（在低频处占主导地位，在结构阻尼极限下按 $\sqrt{4k_{B}T\,I\omega_{0}/Q}$ 比例缩放），$S_{N,\mathrm{ro}}$ 是自准直仪读出贡献（在角度上近似平坦，因此当转换为力矩时按 $\omega^{2}$ 上升，在高频段占主导地位），而 $S_{N,\mathrm{env}}$ 收集了非平稳环境贡献，如地震倾斜、杂散磁场波动以及周围结构的热漂移；已知最后者表现出与测试场地附近人类活动相关的昼夜和每周调制。在驱动线圈不通电的情况下进行的调试运行，在整个测量频带上产生了大约 $10^{-14}\,\mathrm{N\cdot m\,Hz^{-1/2}}$ 数量级的力矩噪声基底，这满足了在第~\ref{sec:level1}节所述的目标精度下验证太极检验质量磁耦合的力矩分辨率要求。
\begin{figure}
% Two-panel figure, to be supplied:
% (top)    time-domain torque segment reconstructed via Eq.~\eqref{eq:dynamics}
% (bottom) amplitude spectral density of the same segment
\includegraphics[width=0.48\textwidth]{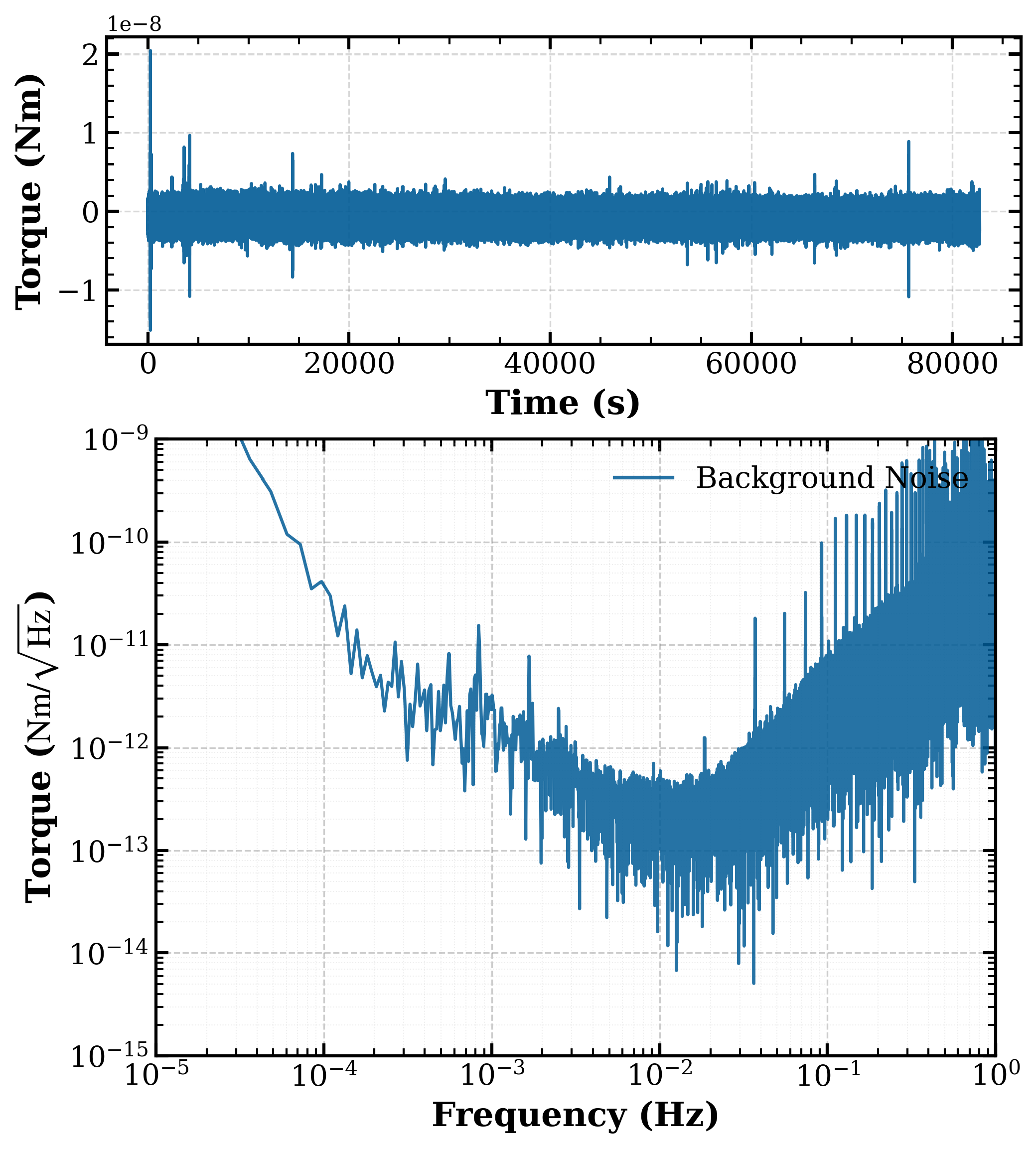}
\caption{\label{fig:raw} Representative raw data from the CIOMP torsion pendulum during commissioning. (Top) Time-domain torque signal reconstructed from the autocollimator angle readout via Eq.~\eqref{eq:dynamics} under free-run conditions, showing slow colored drifts and intermittent transient glitches. (Bottom) Single-sided torque amplitude spectral density of the same run: the low-frequency branch is dominated by environmental drift (rising faster than the structural-damping thermal floor), while the high-frequency branch is consistent with the $\omega^{2}$ rise expected from the autocollimator readout referred to torque.}
%\caption{\label{fig:raw} 调试期间CIOMP扭秤的典型原始数据。（上）自由运行条件下，根据公式~\eqref{eq:dynamics}由自准直仪角度读数重建的时域扭矩信号，显示出缓慢的有色漂移和间歇性的瞬态毛刺。（下）同一次运行的单边扭矩振幅谱密度：低频段由环境漂移主导（其上升速度快于结构阻尼热本底），而高频段与预期中由自准直仪读数折算为扭矩所产生的$\omega^{2}$上升趋势一致。}
\end{figure}

Fig.~\ref{fig:raw} illustrates a representative free-run segment acquired during the CIOMP commissioning campaign. The time-domain panel reveals slow colored drifts and sporadic glitch-like excursions superimposed on the intrinsic random fluctuations; the corresponding ASD panel shows a low-frequency branch steeper than the thermal $1/\sqrt{f}$ prediction (driven by residual environmental drift) and the expected $\omega^{2}$ autocollimator branch at high frequency. The clear non-stationarity and non-whiteness of this background violates the idealized assumptions underpinning standard estimators and directly motivates the data-adaptive weighting strategy introduced in Sec.~\ref{sec:method}.
%图~\ref{fig:raw} 展示了在 CIOMP 设施调试期间获取的一段具有代表性的自由运行数据。时域图揭示了在固有的随机涨落之上，叠加着缓慢的有色漂移和零星的毛刺状突变；对应的振幅谱密度（ASD）图则显示，低频分支比热噪声的 $1/\sqrt{f}$ 预测更为陡峭（由残余环境漂移驱动），高频部分则呈现出预期中的 $\omega^{2}$ 自准直仪分支。该实验背景明显的非平稳性和非白（噪声）特性违背了标准估计器所依赖的理想化假设，这也直接促使我们采用了第~\ref{sec:method}节中介绍的数据自适应加权策略。

\subsection{Magnetic-Coupling Measurement Principle}
\label{sec:principle}
\begin{figure}
\includegraphics[width=0.48\textwidth]{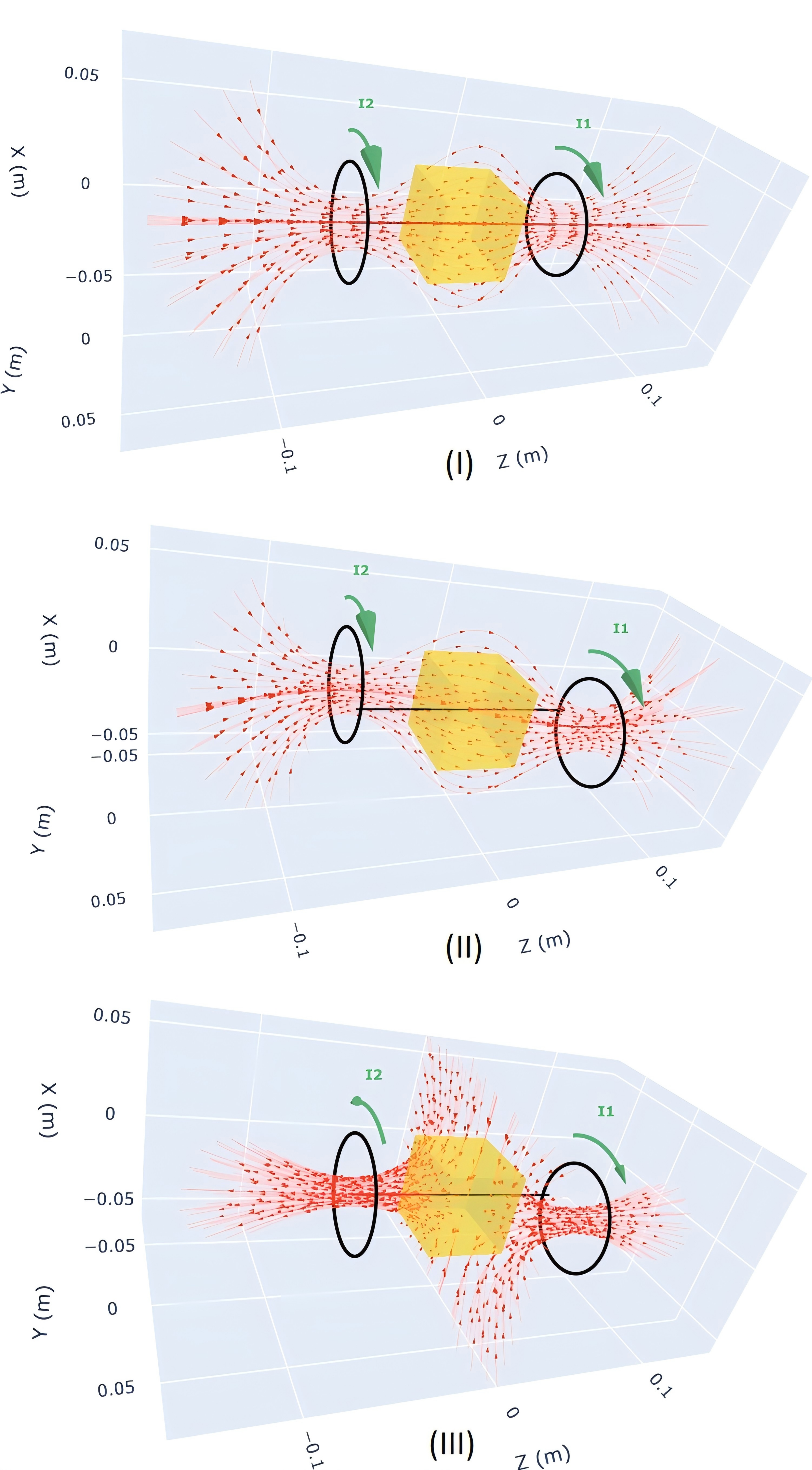}
\caption{\label{fig:configs} Schematic of the three coil configurations employed in the campaign. Red and green arrows indicate the magnetic field lines and current directions, respectively. All coils share identical geometry ($0.03\,\mathrm{m}$ equivalent radius, $240$ turns) and a fixed $z$-axis separation of $0.125\,\mathrm{m}$. Configuration~I: symmetric placement with co-directional currents; Configuration~II: asymmetric placement with co-directional currents; Configuration~III: asymmetric placement with counter-directional currents.}
\end{figure}
%本次实验中采用的三种线圈与电流配置。所有配置的线圈中心在 $z$ 轴上的间距均为 0.125 m，且具有相同的线圈几何形状（等效半径为 0.03 m，240 匝）。配置 I：关于 $xy$ 平面对称放置，电流同向；配置 II：非对称放置于 $(0,\pm 0.025,\pm 0.0625)$ m 处，电流同向；配置 III：非对称放置于 $(0,+0.025,+0.0625)$ m 和 $(0,0,-0.0625)$ m 处，电流反向。

Characterizing the magnetic properties of the TM amounts to reconstructing the components of the remanent magnetic moment $\vec{m}_{r}=(m_{x},m_{y},m_{z})$ and the volume susceptibility $\chi$ from the torque response of the TM to a controlled, time-varying magnetic field applied in its neighborhood. The field is produced by two circular coils flanking the TM along the $z$-direction, driven by independently programmable currents $I_{1}(t)$ and $I_{2}(t)$. By linear superposition, the total applied field is:
%表征检验质量（TM）的磁学性质，归结为根据TM对其附近施加的受控时变磁场的扭矩响应，来重建剩磁矩 $\vec{m}_{r}=(m_{x},m_{y},m_{z})$ 的各个分量以及体积磁化率 $\chi$。该磁场由沿 $z$ 轴方向位于TM两侧的两个圆形线圈产生，由独立可编程的电流 $I_{1}(t)$ 和 $I_{2}(t)$ 驱动。根据线性叠加原理，总的外加磁场为
\begin{equation}
\vec{B}(\vec{r},t) = I_{1}(t)\,\vec{b}_{1}(\vec{r}) + I_{2}(t)\,\vec{b}_{2}(\vec{r}),
\label{eq:B_superposition}
\end{equation}
where $\vec{b}_{j}(\vec{r})$ is the unit-current spatial field of coil $j$, computed from Biot--Savart for the as-built coil geometry.
%中 $\vec{b}{j}(\vec{r})$ 是线圈 $j$ 的单位电流空间磁场，它是根据实际制造的线圈几何结构通过毕奥-萨伐尔定律（Biot-Savart law）计算得出的。

Assuming a spatially uniform remanent magnetization within the TM and zero free current at the TM location (so that $\nabla\times\vec{B}=0$ inside the TM volume), the $x$-axis projection of the total magnetic torque decomposes into a gradient-induced contribution $N_{\mathrm{grad}}$, that arising from the coupling of field gradients with both the remanent and the induced magnetization, and an alignment contribution $N_{m}$ from the cross product of the macroscopic remanent moment with the applied field:
%假设在检验质量（TM）内部具有空间均匀的剩余磁化强度，并且在 TM 所在位置自由电流为零（因此在 TM 体积内 $\nabla\times\vec{B}=0$），总磁力矩在 $x$ 轴上的投影可分解为两个部分：一个是梯度诱导贡献项 $N_{\mathrm{grad}}$，它源于磁场梯度与剩余磁化强度及感生磁化强度的耦合；另一个是取向贡献项 $N_{m}$，它源于宏观剩余磁矩与外加磁场的叉乘：
\begin{equation}
\begin{aligned}
N \;=\;& N_{\mathrm{grad}} + N_{m} \\
 \;=\;& \iiint_{V_{\mathrm{TM}}}\!\Bigg[\, \left( \frac{m_{y}}{V_{\mathrm{TM}}} + \frac{\chi}{\mu_{0}}B_{y} \right)\!\left( y\frac{\partial B_{y}}{\partial z} - z\frac{\partial B_{y}}{\partial y}\right) \\
     & \; + \left( \frac{m_{z}}{V_{\mathrm{TM}}} + \frac{\chi}{\mu_{0}}B_{z} \right)\!\left( y\frac{\partial B_{z}}{\partial z} - z\frac{\partial B_{z}}{\partial y}\right) \\
     & \; + \left( \frac{m_{y}}{V_{\mathrm{TM}}}B_{z} - \frac{m_{z}}{V_{\mathrm{TM}}}B_{y} \right)\Bigg]\,dx\,dy\,dz,
\end{aligned}
\label{eq:torque_integral}
\end{equation}
where $y$ and $z$ are moment arms measured from the TM center. Substituting Eq.~\eqref{eq:B_superposition} into Eq.~\eqref{eq:torque_integral} and factoring the time-varying currents out of the purely geometric spatial integrals, the torque recorded in the $k$-th experimental configuration becomes a linear combination of the three unknowns $(m_{y},m_{z},\chi)$,
%其中 $y$ 和 $z$ 是从TM（检验质量）中心起算的力臂。将公式~\eqref{eq:B_superposition} 代入公式~\eqref{eq:torque_integral} 中，并将随时间变化的电流从纯几何的空间积分中提取出来后，在第 $k$ 种实验配置下记录的扭矩即成为三个未知数 $(m_{y},m_{z},\chi)$ 的线性组合，
\begin{equation}
N^{(k)}(t) = C_{y}^{(k)}(t)\,m_{y} + C_{z}^{(k)}(t)\,m_{z} + C_{\chi}^{(k)}(t)\,\chi,
\label{eq:torque_linear}
\end{equation}
with time-dependent magnetic coupling coefficients
\begin{equation}
\begin{aligned}
C_{y}^{(k)}(t) &= I_{1}^{(k)}(t)\,C_{m_{y},1} + I_{2}^{(k)}(t)\,C_{m_{y},2}, \\
C_{z}^{(k)}(t) &= I_{1}^{(k)}(t)\,C_{m_{z},1} + I_{2}^{(k)}(t)\,C_{m_{z},2}, \\
C_{\chi}^{(k)}(t) &= [I_{1}^{(k)}(t)]^{2}K_{11} + [I_{2}^{(k)}(t)]^{2}K_{22} \\
                  &\quad + I_{1}^{(k)}(t)\,I_{2}^{(k)}(t)\,K_{12},
\end{aligned}
\label{eq:dynamic_coeffs}
\end{equation}
where $C_{m_{i},j}$ and $K_{ij}$ are time-independent tensor integrals of the unit-current fields $\vec{b}_{j}$. Writing $\vec{b}_{j}=(b_{j,x},b_{j,y},b_{j,z})$, the remanence coefficients of the $j$-th coil ($j\in\{1,2\}$) read:
%其中 $C_{m_{i},j}$ 和 $K_{ij}$ 是单位电流磁场 $\vec{b}_{j}$ 的与时间无关的张量积分。记 $\vec{b}_{j}=(b_{j,x},b_{j,y},b_{j,z})$，第 $j$ 个线圈（$j\in\{1,2\}$）的剩磁系数可表示为：
\begin{equation}
\begin{aligned}
C_{m_{y},j} &= \frac{1}{V_{\mathrm{TM}}}\!\iiint_{V_{\mathrm{TM}}}\!\!\left( b_{j,z} + y\frac{\partial b_{j,y}}{\partial z} - z\frac{\partial b_{j,y}}{\partial y}\right)dx\,dy\,dz, \\
C_{m_{z},j} &= \frac{1}{V_{\mathrm{TM}}}\!\iiint_{V_{\mathrm{TM}}}\!\!\left(-b_{j,y} + y\frac{\partial b_{j,z}}{\partial z} - z\frac{\partial b_{j,z}}{\partial y}\right)dx\,dy\,dz,
\end{aligned}
\label{eq:C_coeff}
\end{equation}
while the susceptibility coefficients, quadratic in the fields, are
\begin{equation}
\begin{aligned}
K_{jj} &= \frac{1}{\mu_{0}}\!\iiint_{V_{\mathrm{TM}}}\!\sum_{w\in\{y,z\}} b_{j,w}\!\left( y\frac{\partial b_{j,w}}{\partial z} - z\frac{\partial b_{j,w}}{\partial y}\right)dx\,dy\,dz, \\
K_{12} &= \frac{1}{\mu_{0}}\!\iiint_{V_{\mathrm{TM}}}\!\sum_{w\in\{y,z\}}\!\bigg[\, b_{1,w}\!\left( y\frac{\partial b_{2,w}}{\partial z} - z\frac{\partial b_{2,w}}{\partial y}\right) \\
       &\quad + b_{2,w}\!\left( y\frac{\partial b_{1,w}}{\partial z} - z\frac{\partial b_{1,w}}{\partial y}\right)\bigg] dx\,dy\,dz,
\end{aligned}
\label{eq:K_coeff}
\end{equation}
where the factor of two arising from the cross-term of $|I_{1}\vec{b}_{1}+I_{2}\vec{b}_{2}|^{2}$ has been absorbed into the symmetric definition of $K_{12}$.
%其中，由 $|I_{1}\vec{b}_{1}+I_{2}\vec{b}_{2}|^{2}$ 交叉项产生的系数 2，已被吸收进 $K_{12}$ 的对称定义中

A central feature of Eqs.~\eqref{eq:torque_linear}--\eqref{eq:dynamic_coeffs} is that the remanence and susceptibility channels have different algebraic dependence on the driving currents, linear for $(m_{y},m_{z})$ and quadratic for $\chi$. Driving the coils with pure sinusoids at a common fundamental frequency $f_{\mathrm{mod}}$ therefore physically separates the two channels in the frequency domain: $I(t)\propto\sin(2\pi f_{\mathrm{mod}}t)$ concentrates the remanent-torque response at the fundamental $1\,f_{\mathrm{mod}}$, whereas $I^{2}(t)\propto 1-\cos(4\pi f_{\mathrm{mod}}t)$ up-converts the susceptibility response to a DC term plus the second harmonic $2\,f_{\mathrm{mod}}$. Choosing $f_{\mathrm{mod}}=5\,\mathrm{mHz}$, safely above the pendulum resonance $\omega_{0}/2\pi\simeq 1.85\,\mathrm{mHz}$ yet within the quiet portion of the background spectrum, thus both decouples $(m_{y},m_{z})$ from $\chi$ at the level of the forward model and rejects a large fraction of the low-frequency environmental drift captured by $S_{N,\mathrm{env}}$ in Eq.~\eqref{eq:noise_budget}.
%公式~\eqref{eq:torque_linear}--\eqref{eq:dynamic_coeffs} 的一个核心特征是，剩磁通道和磁化率通道对驱动电流具有不同的代数依赖关系：对于 $(m_{y},m_{z})$ 是线性依赖，对于 $\chi$ 是二次依赖。因此，采用具有共同基频 $f_{\mathrm{mod}}$ 的纯正弦波驱动线圈，即可在频域上从物理层面将这两个通道分离开来：$I(t)\propto\sin(2\pi f_{\mathrm{mod}}t)$ 使剩磁扭矩响应集中在基频 $1\,f_{\mathrm{mod}}$ 处，而 $I^{2}(t)\propto 1-\cos(4\pi f_{\mathrm{mod}}t)$ 则将磁化率响应上变频为一个直流（DC）项加上二次谐波 $2\,f_{\mathrm{mod}}$。选择 $f_{\mathrm{mod}}=5\,\mathrm{mHz}$，该频率安全地高于扭秤的共振频率 $\omega_{0}/2\pi\simeq 1.85\,\mathrm{mHz}$，同时又处于背景噪声谱的安静频段，这不仅在正向模型层面上实现了 $(m_{y},m_{z})$ 与 $\chi$ 的解耦，而且剔除了公式~\eqref{eq:noise_budget} 中 $S_{N,\mathrm{env}}$ 所捕捉到的大部分低频环境漂移。

To break the remaining geometric degeneracies between the unknowns and to provide redundant information for joint estimation, three coil configurations are employed in the present campaign, as illustrated in Fig.~\ref{fig:configs}. In all three, each coil has an equivalent radius of $0.03\,\mathrm{m}$ with $240$ turns, and the projection of the line joining the coil centers onto the $z$-axis is held fixed at $0.125\,\mathrm{m}$; the as-built coil-position tolerance is below $0.3\,\mathrm{mm}$. Configuration~I places the coils symmetrically about the $xy$-plane and drives them co-directionally, maximizing the quasi-uniform field at the TM and therefore the signal-to-noise ratio on the remanent-moment channels $(m_{y},m_{z})$. Configuration~II breaks the $xy$-symmetry by shifting the coil centers to $(0,+0.025,+0.0625)\,\mathrm{m}$ and $(0,-0.025,-0.0625)\,\mathrm{m}$, still co-directional, introducing a controlled field gradient that enhances the susceptibility signal. Configuration~III further de-projects the susceptibility geometric factors by placing the coils at $(0,+0.025,+0.0625)\,\mathrm{m}$ and $(0,0,-0.0625)\,\mathrm{m}$ and driving them counter-directionally. In every run the coils are driven by $f_{\mathrm{mod}}=5\,\mathrm{mHz}$ sinusoids at three different amplitudes, that $0.6$, $0.8$, and $1.0\,\mathrm{A}$, scanning roughly a factor of $1.5$ in signal-to-noise to stress-test estimator robustness across the range of expected campaign conditions. Rotating the TM by $90^\circ$ about the $y$- (or $z$-) axis and repeating the protocol gives access to the remaining remanent component $m_{x}$ via the same formalism. In the present campaign, rotation about the 
$y$-axis was used.
%为了打破未知量之间剩余的几何简并，并为联合估计提供冗余信息，本次实验采用了三种线圈配置，如图 \ref{fig:configs} 所示。在这三种配置中，每个线圈的等效半径均为 0.03 m，匝数为 240 匝，并且连接线圈中心的直线在 $z$ 轴上的投影固定为 0.125 m；建成的线圈位置公差低于 0.3 mm。配置 I 将线圈对称地放置在 $xy$ 平面两侧，并同向驱动它们，从而最大化了 TM（检验质量）处的准均匀场，并因此最大化了剩磁矩通道 $(m_y, m_z)$ 的信噪比。配置 II 打破了 $xy$ 对称性，将线圈中心移动至 (0, +0.025, +0.0625) m 和 (0, -0.025, -0.0625) m 处，依然同向驱动，通过引入受控的场梯度增强了磁化率信号。配置 III 通过将线圈放置在 (0, +0.025, +0.0625) m 和 (0, 0, -0.0625) m 处并反向驱动它们，进一步解耦了磁化率几何因子的投影。在每次运行中，线圈均由频率为 $f_{\mathrm{mod}}$ = 5 mHz 的正弦波驱动，包含 0.8 A、1.0 A 和 1.2 A 三种不同振幅，扫描了大约 1.5 倍的信噪比范围，以便在整个预期实验条件的范围内对估计器的鲁棒性进行压力测试。将 TM 绕 $y$ 轴（或 $z$ 轴）旋转 90° 并重复该实验流程，即可通过相同的数学推导获取剩余的剩磁分量 $m_x$。

\section{The AI-Enhanced Differentiable Weighted Least Squares Framework}
\label{sec:method}

\subsection{From Weighted Least Squares to a Learned Weight Operator}
\label{sec:wls_from_ols}

The forward model derived in Sec.~\ref{sec:principle} establishes that, across any set of $K$ driving configurations and a common acquisition window of $L$ samples, the reconstructed torque time series is a \emph{linear} function of the three unknown magnetic parameters. Stacking the per-configuration torque values $N^{(k)}(t_n)$ row-wise, and filling each row of a design matrix $A\in\mathbb{R}^{M\times 3}$ with the corresponding triple $[\,C_{y}^{(k)}(t_n),\,C_{z}^{(k)}(t_n),\,C_{\chi}^{(k)}(t_n)\,]$ of coupling coefficients from Eq.~\eqref{eq:dynamic_coeffs}, the joint measurement equation across all configurations becomes:
%在第 \ref{sec:principle} 节中推导的正向模型表明，在包含 $K$ 个驱动配置的任意集合以及长度为 $L$ 个样本的公共采集窗口内，重建的扭矩时间序列是三个未知磁参数的\emph{线性}函数。将每个配置的扭矩值 $N^{(k)}(t_n)$ 按行堆叠，并将设计矩阵 $A\in\mathbb{R}^{M\times 3}$ 的每一行填充为等式 \eqref{eq:dynamic_coeffs} 中对应的耦合系数三元组 $[\,C_{y}^{(k)}(t_n),\,C_{z}^{(k)}(t_n),\,C_{\chi}^{(k)}(t_n)\,]$，涵盖所有配置的联合测量方程变为：
\begin{equation}
\mathbf{N} \;=\; A\,\boldsymbol{\beta} \;+\; \mathbf{n}, \qquad \boldsymbol{\beta} \;=\; (m_y,\, m_z,\, \chi)^{\!\top},
\label{eq:linear_model}
\end{equation}
with $\mathbf{N}\in\mathbb{R}^{M}$ the stacked torque vector, $M = KL$ its total length, and $\mathbf{n}\in\mathbb{R}^{M}$ the corresponding stack of noise samples drawn from the pendulum background characterized in Sec.~\ref{sec:pendulum}. If $\mathbf{n}$ were zero-mean with covariance $\Sigma$, the Gauss--Markov theorem would identify the best linear unbiased estimator of $\boldsymbol{\beta}$ as the generalized least-squares solution $\hat{\boldsymbol{\beta}}_{\mathrm{GLS}} = (A^{\!\top}\Sigma^{-1}A)^{-1}A^{\!\top}\Sigma^{-1}\mathbf{N}$; in the diagonal-covariance limit this reduces to weighted least squares (WLS) with weights $w_n = 1/\sigma_n^{2}$, and in the further white-noise limit $\sigma_n = \mathrm{const}$ it degenerates to OLS.
%其中 $\mathbf{N}\in\mathbb{R}^{M}$ 为堆叠后的扭矩向量，$M = KL$ 为其总长度，$\mathbf{n}\in\mathbb{R}^{M}$ 为从第 \ref{sec:pendulum} 节中表征的扭摆背景中提取的对应噪声样本堆叠。如果 $\mathbf{n}$ 满足零均值且协方差为 $\Sigma$，根据高斯-马尔可夫定理（Gauss--Markov theorem），$\boldsymbol{\beta}$ 的最佳线性无偏估计量即为广义最小二乘解 $\hat{\boldsymbol{\beta}}_{\mathrm{GLS}} = (A^{\!\top}\Sigma^{-1}A)^{-1}A^{\!\top}\Sigma^{-1}\mathbf{N}$；在对角协方差极限情况下，该解简化为权重为 $w_n = 1/\sigma_n^{2}$ 的加权最小二乘法（WLS）；在进一步的白噪声极限条件 $\sigma_n = \mathrm{const}$ 下，它则退化为普通最小二乘法（OLS）。

In realistic torsion-pendulum operation, however, the noise fails all three premises: Eq.~\eqref{eq:noise_budget} is colored in frequency, the environmental contribution $S_{N,\mathrm{env}}$ is non-stationary on diurnal and weekly time scales, and the raw stream is additionally contaminated by localized transients (seismic micro-events, air-conditioning glitches, autocollimator dropouts) whose sample-by-sample variance cannot be captured by any stationary $\Sigma$. The consequence, quantified in Sec.~\ref{sec:results}, is that both OLS and a KF with a stationary-noise model develop systematic biases in $\hat{\chi}$ at a level comparable to or exceeding the Taiji ground-test target.
%然而，在真实的扭摆运行中，噪声违背了所有三个前提：等式 \eqref{eq:noise_budget} 在频域上表现为有色噪声，环境带来的贡献 $S_{N,\mathrm{env}}$ 在昼夜和周的时间尺度上是非平稳的，而且原始数据流还受到局部瞬态事件（微小地震事件、空调故障、自准直仪信号丢失）的额外污染，其逐样本方差无法被任何平稳的协方差矩阵 $\Sigma$ 所捕获。其后果（如第 \ref{sec:results} 节所量化）是，普通最小二乘法 (OLS) 和基于平稳噪声模型的卡尔曼滤波器均会在 $\hat{\chi}$ 中产生系统性偏差，其量级与太极计划地面测试目标相当甚至超出了该目标。

The key observation motivating the present work is that the \emph{optimal} diagonal weight $w_n$ at sample $t_n$ is a function of the local noise state, not of a single global covariance; if this local state can be inferred directly from the data stream, the WLS estimator recovers its Gauss--Markov optimality in a running, sample-local sense. We therefore replace the hand-designed weight vector by a learned operator,
\begin{equation}
\mathbf{w} \;=\; f_{\boldsymbol{\phi}}(\mathbf{N}),\qquad \mathbf{w}\in\mathbb{R}^{M}_{+},
\label{eq:weight_operator}
\end{equation}
implemented as a neural network $f_{\boldsymbol{\phi}}$ with trainable parameters $\boldsymbol{\phi}$. In practice the network is applied to one configuration stream $N^{(k)}(t)$ at a time, producing per-configuration weights $w^{(k)}(t)$; these are concatenated into the global diagonal $W = \mathrm{diag}(\mathbf{w})$ for the joint solve described below. The network emits a strictly positive per-sample weight that is embedded in a fully differentiable WLS solver; the parameter-estimation error on $\hat{\boldsymbol{\beta}}$ is then backpropagated through the solver into $\boldsymbol{\phi}$, allowing the network to learn, end-to-end, what "contaminated" means \emph{for the specific downstream task of recovering} $(m_y, m_z, \chi)$ rather than for a generic denoising objective. This single design choice distinguishes AI-WLS from both classical WLS (where $\mathbf{w}$ is fixed by an offline PSD estimate) and from "black-box" regression approaches (where a network predicts $\boldsymbol{\beta}$ directly, discarding the known linear physics of Eq.~\eqref{eq:linear_model}).
%促成本项研究的关键观察在于，样本 $t_n$ 处的\emph{最优}对角权重 $w_n$ 是局部噪声状态的函数，而不是单一全局协方差的函数；如果可以直接从数据流中推断出这种局部状态，那么 WLS（加权最小二乘）估计器就能在一种连续的、样本局部的意义上恢复其高斯-马尔可夫最优性。因此，我们将人工设计的权重向量替换为一个通过学习得到的算子：$$\mathbf{w} \;=\; f_{\boldsymbol{\phi}}(\mathbf{N}),\qquad \mathbf{w}\in\mathbb{R}^{M}_{+}, \label{eq:weight_operator}$$该算子被实现为具有可训练参数 $\boldsymbol{\phi}$ 的神经网络 $f_{\boldsymbol{\phi}}$。在实际操作中，该网络每次应用于单个配置的数据流 $N^{(k)}(t)$，生成对应每个配置的权重 $w^{(k)}(t)$；这些权重被拼接成全局对角矩阵 $W = \mathrm{diag}(\mathbf{w})$，用于下文所述的联合求解。该网络输出严格为正的逐样本权重，并将其嵌入到一个完全可微的 WLS 求解器中；随后，$\hat{\boldsymbol{\beta}}$ 上的参数估计误差会通过求解器反向传播至 $\boldsymbol{\phi}$，从而使网络能够端到端地学习所谓“被污染”\emph{对于恢复 $(m_y, m_z, \chi)$ 这一特定下游任务}的真实含义，而非仅仅针对一个通用的去噪目标。这一独特的设计选择使得 AI-WLS 既区别于经典的 WLS（在经典方法中 $\mathbf{w}$ 由离线 PSD 估计固定），也区别于“黑盒”回归方法（在这种方法中，网络直接预测 $\boldsymbol{\beta}$，从而抛弃了等式 \eqref{eq:linear_model} 中已知的线性物理规律）。

\begin{figure*}
\includegraphics[width=0.99\textwidth]{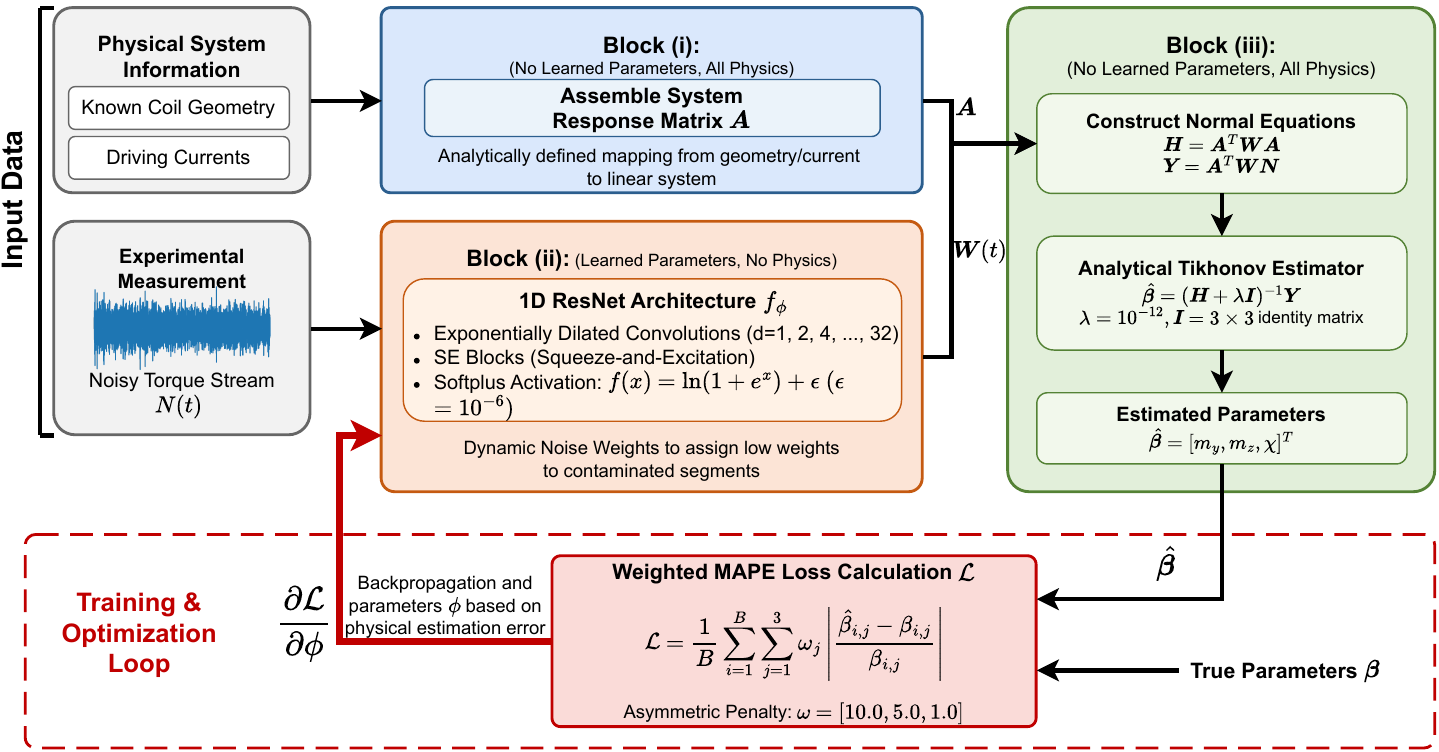}
\caption{\label{fig:pipeline} Schematic of the AI-WLS pipeline. The measured torque stream $\mathbf{N}$ is consumed by the dilated ResNet noise evaluator, which emits a strictly positive per-sample weight $\mathbf{w}$. These weights are combined with the physics-fixed design matrix $A$ (assembled from the known coil geometry and driving currents) inside a differentiable weighted least-squares solver that returns the parameter vector $\hat{\boldsymbol{\beta}}=(m_y,m_z,\chi)^{\top}$. The parameter-estimation error flows back end-to-end through the solver to update the network weights $\boldsymbol{\phi}$.The red dashed box encloses the loss function during the training phase.}
\end{figure*}

The overall pipeline, summarized in Fig.~\ref{fig:pipeline}, factorizes cleanly into three blocks: (i) a physics-fixed forward operator that assembles $A$ from the known coil geometry and driving currents; (ii) a learned noise-evaluator network $f_{\boldsymbol{\phi}}$ that consumes the measured torque stream and emits per-sample weights; and (iii) a differentiable WLS solver that returns $\hat{\boldsymbol{\beta}}$ in closed form. Blocks (i) and (iii) carry all of the physics and none of the learned parameters; block (ii) carries all of the learned parameters and none of the physics. Training and inference therefore never confuse signal modeling with noise modeling.
%总体流程（如图 \ref{fig:pipeline} 总结所示）清晰地划分为三个模块：(i) 一个固定物理模型的前向算子，根据已知的线圈几何结构和驱动电流构建矩阵 $A$；(ii) 一个通过学习得到的噪声评估网络 $f_{\boldsymbol{\phi}}$，用于接收测量到的扭矩数据流并输出逐样本权重；以及 (iii) 一个可微的 WLS（加权最小二乘）求解器，以闭式解的形式返回 $\hat{\boldsymbol{\beta}}$。模块 (i) 和 (iii) 包含了所有的物理机制，且不包含任何可学习参数；模块 (ii) 包含了所有的可学习参数，且不涉及任何物理机制。因此，无论是训练还是推理过程，都绝不会将信号建模与噪声建模混为一谈。

%AI-WLS 流程示意图。测量所得的扭矩数据流 $\mathbf{N}$ 输入至空洞 ResNet（Dilated ResNet）噪声评估器，该评估器输出严格为正的逐样本权重 $\mathbf{w}$。这些权重与固定物理模型的设计矩阵 $A$（由已知的线圈几何结构和驱动电流构建）在一个可微的加权最小二乘求解器中相结合，该求解器返回参数向量 $\hat{\boldsymbol{\beta}}=(m_y,m_z,\chi)^{\top}$。参数估计误差通过求解器进行端到端的反向传播，以更新网络权重 $\boldsymbol{\phi}$。}

\subsection{Dilated-Residual Noise Evaluator with Squeeze--Excitation Recalibration}
\label{sec:resnet}
 
Two physical features of the problem constrain the architecture of $f_{\boldsymbol{\phi}}$. First, the dominant non-idealities of the torsion-pendulum background are localized in time, that transient glitches (autocollimator dropouts, seismic micro-events, occasional HVAC-induced excursions) occupy from sub-second to several-second spans, while the broader coloring of $S_{N,\mathrm{env}}$ is already substantially rejected at the forward-model level by the AC modulation scheme of Sec.~\ref{sec:principle}. Second, these transients can sit on top of, or interfere constructively with, the driven sinusoid at $f_\mathrm{mod} = 5\,\mathrm{mHz}$, so the network must distinguish genuine signal modulation from excursions while operating on short local neighborhoods. The task of the noise evaluator is therefore explicitly local anomaly detection, not global drift tracking, that the latter is handled by the physics of the modulation and solver.
%该问题的两个物理特征约束了 $f_{\boldsymbol{\phi}}$ 的架构。首先，扭摆背景中主要的非理想特性具有时间局部性，即瞬态毛刺（自准直仪信号丢失、微型地震事件、偶发的空调引起的波动）的持续时间在亚秒到几秒之间，而 $S_{N,\mathrm{env}}$ 更宽泛的有色噪声特征，则已经在正向模型层面被第 \ref{sec:principle} 节所述的交流调制方案很大程度上抑制了。其次，这些瞬态扰动可能会叠加在 $f_\mathrm{mod} = 5\,\mathrm{mHz}$ 的受驱正弦波上，或者与之发生相长干涉，因此，网络必须在基于短时间局部邻域进行操作时，能够将真实的信号调制与异常波动区分开来。因此，噪声评估器的任务明确为局部异常检测，而非全局漂移跟踪，因为后者已经由调制机制和求解器本身的物理属性处理了。

We implement $f_{\boldsymbol{\phi}}$ as a one-dimensional residual network built from exponentially dilated convolutions. Each residual block applies a causal dilated convolution with kernel size $3$ and dilation rate $d_\ell$, followed by a nonlinearity and a skip connection; stacking six such blocks with $d_\ell = 1, 2, 4, 8, 16, 32$ yields an effective receptive field of
\begin{equation}
R_{\mathrm{eff}} \;=\; 1 + 2\sum_{\ell=1}^{6} d_\ell \;=\; 127 \;\text{samples},
\label{eq:rf}
\end{equation}
or roughly $12.7\,\mathrm{s}$ at the $10\,\mathrm{Hz}$ acquisition rate. This is deliberately shorter than one $200\,\mathrm{s}$ driving period: each output weight $w_n$ is informed only by a local neighborhood of its own sample, in line with the anomaly-detection role of the network.
%我们将 $f_{\boldsymbol{\phi}}$ 实现为由指数级空洞卷积（dilated convolutions）构建的一维残差网络。每个残差块应用一个卷积核大小为 3、空洞率为 $d_\ell$ 的因果空洞卷积，其后紧跟一个非线性激活函数和一个跳跃连接；将 $d_\ell = 1, 2, 4, 8, 16, 32$ 的六个这样的模块堆叠起来，可得到如下的有效感受野：$$R_{\mathrm{eff}} \;=\; 1 + 2\sum_{\ell=1}^{6} d_\ell \;=\; 127 \;\text{ 个样本}, \label{eq:rf}$$在 $10\,\mathrm{Hz}$ 的采集速率下，这大约相当于 $12.7\,\mathrm{s}$。这一时间跨度被刻意设计成短于 $200\,\mathrm{s}$ 的单次驱动周期：每个输出权重 $w_n$ 的信息仅来源于其所属样本的局部邻域，这与该网络所承担的异常检测角色相契合。

Within each residual block we insert a Squeeze-and-Excitation (SE) module~\cite{Hu2018SENet} that performs channel-wise feature recalibration: a global average pooling across the temporal axis of the block produces a per-channel descriptor, which is passed through a two-layer bottleneck and a sigmoid gate to yield multiplicative channel weights. The SE module is the mechanism by which the network selectively amplifies channels that have locked onto transient-glitch morphology while suppressing channels that respond only to the clean driven sinusoid (which must \emph{not} be down-weighted, since it carries the signal).
%在每个残差块中，我们插入了一个 Squeeze-and-Excitation (SE) 模块~\cite{Hu2018SENet}，用于执行通道级的特征重校准：沿着该模块的时间轴进行全局平均池化会生成一个逐通道描述符，该描述符随后穿过一个双层瓶颈结构和一个 sigmoid 门控，从而产生乘性的通道权重。正是通过 SE 模块这一机制，网络能够选择性地放大那些已捕捉到瞬态毛刺（glitch）形态的通道，同时抑制那些仅对纯净受驱正弦波产生响应的通道（由于后者承载着信号，因此\emph{绝不能}被降低权重）。

The output head is a $1\times 1$ convolution to a single scalar channel followed by a Softplus activation with a small positive floor,
\begin{equation}
w_n \;=\; \mathrm{Softplus}\big(z_n\big) + \varepsilon \;=\; \ln\!\big(1 + e^{z_n}\big) + \varepsilon,
\label{eq:softplus}
\end{equation}
with $\varepsilon = 10^{-6}$ and $z_n$ the pre-activation. Two properties of Eq.~\eqref{eq:softplus} are essential. (i)~Strict positivity, $w_n > 0$, guarantees that the weighted normal matrix introduced below is positive-definite and the WLS solve is well-posed for every batch and every epoch; a ReLU output would fail this as soon as one pre-activation turned negative, and a sigmoid would saturate. (ii)~Smoothness, including smoothness across $z = 0$, preserves gradient flow through the network at all training steps, which is critical because the learning signal reaches $\boldsymbol{\phi}$ only through the downstream solver.
%输出头是一个 $1\times 1$ 卷积，输出至单个标量通道，随后是一个带有微小正下限的 Softplus 激活函数：$$w_n \;=\; \mathrm{Softplus}\big(z_n\big) + \varepsilon \;=\; \ln\!\big(1 + e^{z_n}\big) + \varepsilon, \label{eq:softplus}$$其中 $\varepsilon = 10^{-6}$，$z_n$ 为预激活值（pre-activation）。等式 \eqref{eq:softplus} 的两个属性至关重要：(i) 严格为正（$w_n > 0$）：这保证了下文引入的加权正规矩阵（weighted normal matrix）是正定的，并且加权最小二乘（WLS）的求解在每个批次（batch）和每个训练轮次（epoch）中都是适定的；如果是 ReLU 输出，一旦某个预激活值变为负数就会破坏这一属性，而 sigmoid 输出则会导致梯度饱和。(ii) 平滑性（包括在 $z = 0$ 处的平滑性）：这能在所有的训练步骤中维持梯度在整个网络中的流动，这一点极其关键，因为学习信号仅仅只能通过下游的求解器才能传递到达 $\boldsymbol{\phi}$。

\subsection{Differentiable WLS Solver and Loss Function}
\label{sec:solver_loss}

Given a weight vector $\mathbf{w}$ emitted by the network and the physics-fixed design matrix $A$, the differentiable WLS solver performs, in closed form,
\begin{equation}
\hat{\boldsymbol{\beta}} \;=\; \big(A^{\!\top}W A + \lambda \mathbb{I}_{3}\big)^{-1} A^{\!\top} W \mathbf{N},
\label{eq:wls_solve}
\end{equation}
where $W = \mathrm{diag}(\mathbf{w})$, $\mathbb{I}_{3}$ is the $3\times 3$ identity, and $\lambda = 10^{-12}$ is a Tikhonov term included purely for numerical conditioning. Here the physical scale of $A^{\!\top}W A$ is many orders of magnitude larger than $\lambda$, so the regularization does not bias the estimate, but it prevents catastrophic pivot failures in the early training epochs when $\mathbf{w}$ has not yet organized itself. In the limit $\mathbf{w}\to\mathbf{1}$ and $\lambda\to 0$, Eq.~\eqref{eq:wls_solve} reduces to the OLS estimator used as a baseline in Sec.~\ref{sec:results}, so AI-WLS contains OLS as a strict special case and can never be asymptotically worse in expectation provided the network is allowed to reach that fixed point.
%给定由网络输出的权重向量 $\mathbf{w}$ 以及固定物理模型的设计矩阵 $A$，可微的 WLS（加权最小二乘）求解器以闭式解的形式执行如下计算：$$\hat{\boldsymbol{\beta}} \;=\; \big(A^{\!\top}W A + \lambda \mathbb{I}_{3}\big)^{-1} A^{\!\top} W \mathbf{N}, \label{eq:wls_solve}$$其中 $W = \mathrm{diag}(\mathbf{w})$，$\mathbb{I}_{3}$ 为 $3\times 3$ 单位矩阵，$\lambda = 10^{-12}$ 是纯粹为了改善数值条件而引入的吉洪诺夫（Tikhonov）正则化项。此处 $A^{\!\top}W A$ 的物理尺度比 $\lambda$ 大许多个数量级，因此该正则化不会使估计产生偏差，但它能防止在训练早期的轮次（epoch）中，由于 $\mathbf{w}$ 尚未收敛成型而导致的灾难性主元失效（pivot failures）。在 $\mathbf{w}\to\mathbf{1}$ 且 $\lambda\to 0$ 的极限情况下，等式 \eqref{eq:wls_solve} 退化为第 \ref{sec:results} 节中作为基准使用的 OLS（普通最小二乘）估计器；因此，AI-WLS 将 OLS 作为一个严格的特例包含在内，并且只要允许网络达到该不动点，其在期望上的渐近表现就绝不会更差。

The solve is implemented via a Cholesky factorization of the $3\times 3$ weighted normal matrix; both the factorization and the triangular back-substitution have closed-form derivatives that are native to modern automatic-differentiation frameworks, so $\partial\hat{\boldsymbol{\beta}}/\partial\mathbf{w}$ and hence $\partial\hat{\boldsymbol{\beta}}/\partial\boldsymbol{\phi}$ are obtained without numerical differentiation. The small size of the solve ($3\times 3$) makes the solver effectively free compared with the forward pass through $f_{\boldsymbol{\phi}}$, so batch sizes and window lengths are unconstrained by the solver itself.
%该求解过程是通过对 $3\times 3$ 加权正规矩阵进行 Cholesky 分解（乔列斯基分解）来实现的；无论是分解过程还是三角回代均具有闭式导数，且现代自动微分框架对此具有原生支持。因此，无需进行数值微分即可求得 $\partial\hat{\boldsymbol{\beta}}/\partial\mathbf{w}$，并进一步得到 $\partial\hat{\boldsymbol{\beta}}/\partial\boldsymbol{\phi}$。由于求解规模很小（$3\times 3$），与通过 $f_{\boldsymbol{\phi}}$ 的前向传播相比，求解器的计算成本几乎可以忽略不计，因此批次大小（batch sizes）和窗口长度完全不受求解器本身的限制。

A note on gauge. For any constant $c>0$, the rescaling $\mathbf{w}\mapsto c\,\mathbf{w}$ sends the solve to $\hat{\boldsymbol{\beta}} = (A^{\!\top}W A + (\lambda/c)\mathbb{I}_3)^{-1} A^{\!\top} W \mathbf{N}$, i.e. it leaves the estimate invariant up to a rescaling of the Tikhonov term; given the very small $\lambda$ adopted here, the estimate is effectively invariant under $c$ and only the \emph{relative} weight profile across samples affects $\hat{\boldsymbol{\beta}}$. This gauge freedom is resolved implicitly by the Softplus output and the finite Tikhonov floor; no explicit normalization of $\mathbf{w}$ is imposed.
%对于任何常数 $c>0$，缩放变换 $\mathbf{w}\mapsto c\,\mathbf{w}$ 会使求解结果变为 $\hat{\boldsymbol{\beta}} = (A^{\!\top}W A + (\lambda/c)\mathbb{I}_3)^{-1} A^{\!\top} W \mathbf{N}$。也就是说，这种缩放只会导致 Tikhonov 正则化项发生相应的比例变化，而保持估计值基本不变。鉴于本文所采用的 $\lambda$ 值极小，估计值对 $c$ 的变化实际上是不敏感的（即具有不变性），只有样本间的相对权重分布才会影响 $\hat{\boldsymbol{\beta}}$。这种权重缩放的自由度（Gauge freedom）已通过 Softplus 输出层和设定的 Tikhonov 正则化下限隐式地解决了，因此无需对 $\mathbf{w}$ 进行显式的归一化处理。

Because the three components of $\boldsymbol{\beta}$ span several orders of magnitude ($m_{y,z}\sim 10^{-8}\,\mathrm{A\cdot m^{2}}$ versus $\chi\sim 10^{-5}$) a loss function expressed in absolute error would be numerically dominated by whichever component happens to be largest at initialization and would fail to constrain the others. We therefore adopt a weighted mean-absolute-percentage-error (WMAPE) loss,
\begin{equation}
\mathcal{L}(\boldsymbol{\phi}) \;=\; \frac{1}{B}\sum_{i=1}^{B}\sum_{j=1}^{3} \omega_{j}\,\Bigg|\frac{\hat{\beta}_{i,j} - \beta_{i,j}^{\mathrm{true}}}{\beta_{i,j}^{\mathrm{true}}}\Bigg|,
\label{eq:wmape}
\end{equation}
where $B$ is the mini-batch size and the index pair $(i,j)$ runs over batch element and component of $\boldsymbol{\beta}=(m_y, m_z, \chi)^{\!\top}$ respectively. Relative error is the natural objective for parameters whose scientific requirement is itself stated as a fractional precision, and WMAPE is robust to the heavy-tailed statistics of the driven-signal estimator (which has occasional outliers dominated by glitch-contaminated batches) in a way that a squared loss is not.

The per-component penalty vector $\boldsymbol{\omega}=(\omega_1,\omega_2,\omega_3)$, with components corresponding respectively to $m_y$, $m_z$, and $\chi$, encodes a physics priority. Because the magnetic-noise budget of Taiji is more sensitive to $|\vec{m}_{r}|$ than to $\chi$ under the expected ambient-field spectrum, and because the fractional precision required on the dipole moments is more stringent than on the susceptibility (by comparison of the targets $10^{-9}\,\mathrm{A\cdot m^{2}}$ vs $10^{-7}$ against the expected magnitudes), we adopt $\boldsymbol{\omega} = (10.0,\,5.0,\,1.0)$ in all reported runs. 
Empirical results from traditional estimation methods indicate substantially larger errors for the $m_y$ and $m_z$ components. We therefore assign higher weights to these two parameters to ensure they are sufficiently constrained during training.
% The exact numerical value of this prior is not critical, that we observed empirically that values in the range $\omega_1/\omega_3=\omega_2/\omega_3\in[3,10]$ produce statistically indistinguishable test-set results, with degraded dipole precision for ratios $\lesssim 1$ and degraded $\chi$ precision for ratios $\gtrsim 20$, but it must be non-trivial, because the uniform choice $\boldsymbol{\omega} = (1,1,1)$ converges to a solution biased toward the (larger-relative-gradient) $\chi$ channel.

\subsection{Training Protocol and Data Handling}
\label{sec:training}

Training pairs are assembled by superposing physically consistent simulated signals $\mathbf{N}_{\mathrm{clean}} = A\boldsymbol{\beta}^{\mathrm{true}}$ onto real, uncalibrated background-noise segments extracted from the CIOMP commissioning runs described in Sec.~\ref{sec:pendulum}. Ground-truth parameters $\boldsymbol{\beta}^{\mathrm{true}}$ are drawn from the physically motivated prior
\begin{equation}
m_x,\, m_y,\,m_z\in[-10,\,10]\times 10^{-8}\,\mathrm{A\cdot m^{2}},\quad \chi\in[1,\,10]\times 10^{-5},
\end{equation}
with independent uniform signs, comfortably bracketing the Taiji expected values while keeping the error denominators bounded away from zero; the corresponding $A\boldsymbol{\beta}^{\mathrm{true}}$ templates are recomputed on-the-fly using the actual driving-current waveforms, so the network is never exposed to a cached signal library.

The handling of noise realizations is the critical step for generalization. 
To faithfully represent actual detector conditions, continuous segments of 10,000 data points are randomly extracted from the background-noise recording to serve as realistic noise realizations.
During training only, we apply two augmentations to the training-set noise segments before the signal is superposed: (a) random continuous slicing from the chronological noise pool, which breaks any accidental phase coherence between glitch patterns and the driving sinusoid in the training distribution; and (b) amplitude jittering by a uniform factor $\in[0.8, 1.2]$, which regularizes the network against overfitting to the specific mean noise level of the commissioning run. 
%噪声实现的处理是实现泛化能力的关键步骤。为了真实反映实际的探测器条件，从背景噪声记录中随机提取长度为 10,000 个数据点的连续片段，作为真实的噪声实现。
%仅在训练阶段，我们在叠加信号之前对训练集噪声片段应用两种增强方法：(a) 从按时间顺序排列的噪声池中随机连续切片，这可以打破训练分布中毛刺模式与驱动正弦波之间任何偶然的相位相干性；(b) 乘以一个服从均匀分布 $[0.8, 1.2]$ 的幅度抖动因子，这可以对网络进行正则化，防止其过度拟合调试运行期间特定的平均噪声水平。

The network is trained with the AdamW optimizer (weight decay factor of $10^{-3}$) at an initial learning rate of $5 \times 10^{-5}$ on mini-batches of $B = 32$ windows of length $L = 10{,}000$ samples ($1000\,\mathrm{s}$, i.e. five full modulation periods). We employ a cosine annealing learning rate schedule over a maximum budget of $1500$ epochs, automatically saving the model checkpoint that achieves the lowest validation loss. All reported numerical results (Sec.~\ref{sec:results}) are evaluated on the chronologically held-out test-set noise that the network has never seen, with templates that use driving-current amplitudes and configurations matching the physical experiment. Importantly, the OLS and KF baselines against which AI-WLS is benchmarked are evaluated on the \emph{same} signal-plus-noise realizations as AI-WLS, so any performance difference reflects only the estimator, not an input-data discrepancy.
%网络使用 AdamW 优化器（权重衰减因子为 $10^{-3}$）进行训练，初始学习率为 $5 \times 10^{-5}$，小批量大小为 $B = 32$ 个窗口，每个窗口长度为 $L = 10{,}000$ 个样本（即 $1000,\mathrm{s}$，相当于五个完整的调制周期）。我们采用余弦退火学习率调度，最大训练轮数为 $1500$ 轮，并自动保存达到最低验证损失的模型检查点。所有报告的数值结果（第~\ref{sec:results} 节）均在按时间顺序保留的、网络从未见过的测试集噪声上进行评估，所使用的模板与物理实验中的驱动电流幅值和构型相匹配。重要的是，用于与 AI-WLS 进行基准比较的 OLS 和 KF 基线，是在与 AI-WLS 相同的信号加噪声实现上进行评估的，因此任何性能差异仅反映估计器本身，而非输入数据的差异。

\section{Experiments and Results}
\label{sec:results}

\subsection{Injection Protocol}
\label{sec:injection}

As anticipated in Sec.~\ref{sec:pendulum}, the present campaign uses the CIOMP torsion pendulum in its commissioning state. A calibrated physical reference dipole of known $(\vec{m}_r,\chi)$ is not yet available, so we validate AI-WLS through controlled signal injection: physically consistent magnetic torque templates are superposed onto the real, uncalibrated background noise recorded by the pendulum, and the three estimators (AI-WLS, OLS, and KF) are benchmarked on the same signal plus noise realizations. The ground truth in this protocol is the injected $\boldsymbol{\beta}^{\mathrm{true}}=(m_y, m_z, \chi)^{\!\top}$, and any estimator bias or variance reflects only the interplay of the colored, non-stationary background with the estimator itself, not an input data discrepancy.
%正如第~\ref{sec:pendulum}~节所预期的那样，本次实验使用的是处于调试阶段的 CIOMP 扭摆。由于目前尚未配备已知 $(\vec{m}_r,\chi)$ 且经过校准的物理参考偶极子，因此我们通过受控的信号注入来验证 AI-WLS：将符合物理规律的磁扭矩模板叠加到扭摆记录的真实、未校准的背景噪声上，并在相同的“信号加噪声”样本上对三种估计器（AI-WLS、OLS 和卡尔曼滤波器）进行基准测试。在该方案中，真实值（ground truth）即为注入的 $\boldsymbol{\beta}^{\mathrm{true}}=(m_y, m_z, \chi)^{\!\top}$，任何估计器的偏差或方差都仅反映了有色、非平稳背景与估计器本身之间的相互影响，而并非输入数据存在差异。

The injected templates are generated from the forward model of Eqs.~\eqref{eq:torque_linear} and \eqref{eq:dynamic_coeffs}, using the as built coil geometry of Fig.~\ref{fig:configs} (equivalent radius $0.03\,\mathrm{m}$, $240$ turns, $z$ axis coil spacing $0.125\,\mathrm{m}$, coil position tolerance below $0.3\,\mathrm{mm}$) and the sinusoidal driving currents specified in Sec.~\ref{sec:principle}, at the fundamental frequency $f_{\mathrm{mod}}=5\,\mathrm{mHz}$ and the three campaign amplitudes $\{0.6,\,0.8,\,1.0\}\,\mathrm{A}$. Within each run, the three coil configurations are engaged in sequence. Furthermore, determining the $m_x$ component necessitates reorienting the TM and repeating the full experimental protocol, which generates a total of 18 testing scenarios. The ground truth values are fixed at  $m_x = -9.0230\times 10^{-8}\,\mathrm{A\cdot m^{2}}$,$m_y = -3.1700\times 10^{-8}\,\mathrm{A\cdot m^{2}}$, $m_z = -9.8120\times 10^{-8}\,\mathrm{A\cdot m^{2}}$, and $\chi = 9.0900\times 10^{-5}$, values representative of a quiet copper tungsten TM and comfortably within the training prior of Sec.~\ref{sec:training}.
%注入的模板是基于公式~\eqref{eq:torque_linear}和\eqref{eq:dynamic_coeffs}的正向模型生成的，其中使用了图~\ref{fig:configs}中实际搭建的线圈几何结构（等效半径 0.03 m，240 匝，z 轴线圈间距 0.125 m，线圈位置公差小于 0.3 mm），以及第~\ref{sec:principle}节中规定的正弦驱动电流，其基频为 $f_{\mathrm{mod}}$ = 5 mHz，且包含三种实验测试幅值 {0.6, 0.8, 1.0} A。在每次运行中，三种线圈配置会被依次投入测试。并且，为了标定 $m_x$ 分量，需要调整TM的方向并重复整个实验流程，从而共产生 18 种测试场景。各项参数的真值固定为 $m_x = -9.0230\times 10^{-8}\,\mathrm{A\cdot m^{2}}$、$m_y = -3.1700\times 10^{-8}\,\mathrm{A\cdot m^{2}}$、$m_z = -9.8120\times 10^{-8}\,\mathrm{A\cdot m^{2}}$，以及 $\chi = 9.0900\times 10^{-5}$，这些数值是典型低磁态铜钨检验质量 (TM) 的代表值，并且完全落在了第~\ref{sec:training}节设定的训练先验范围之内。

\begin{figure}
\includegraphics[width=0.45\textwidth]{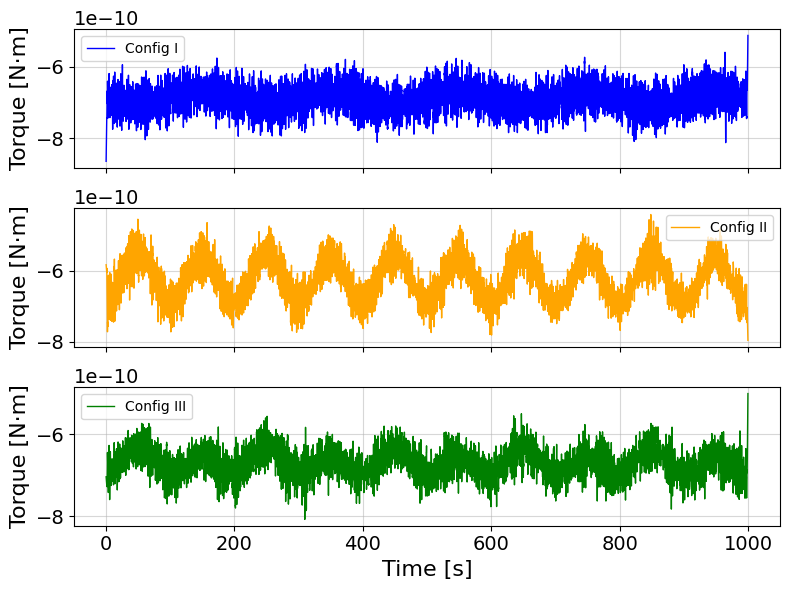}
\caption{\label{fig:angles} Representative $1000\,\mathrm{s}$ segments of the simulated TM torque $N(t)$ driven by a $0.6\,\mathrm{A}$ current. The panels correspond to the three coil configurations depicted in Fig.~\ref{fig:configs}: Configuration~I (top), Configuration~II (middle), and Configuration~III (bottom). }
\end{figure}

Fig.~\ref{fig:angles} displays a representative $1000\,\mathrm{s}$ segment of the simulated TM torque under $0.6\,\mathrm{A}$ driving, one segment per configuration. The signature predicted by Eq.~\eqref{eq:dynamic_coeffs} is clearly visible by eye: Configuration~I, with the coils arranged symmetrically and driven co directionally, produces a response dominated by the fundamental at $1\,f_{\mathrm{mod}}$ and therefore carries most of the information on the remanent moment components $(m_y,m_z)$; Configurations~II and III, which introduce a controlled geometric asymmetry, are dominated by the second harmonic $2\,f_{\mathrm{mod}}$ and thereby amplify the susceptibility channel $\chi$. The combination of the three configurations therefore breaks the column degeneracies of the design matrix $A$ and provides the redundancy needed for the three parameter joint fit performed in Sec.~\ref{sec:method}.
%图~\ref{fig:angles} 展示了在 $1.2\,\mathrm{A}$ 驱动下具有代表性的 $1000\,\mathrm{s}$ 模拟 TM 角位移片段，每种配置对应一个片段。肉眼可以清晰地看到等式~\eqref{eq:dynamic_coeffs} 所预测的信号特征：配置 I 的线圈对称排列且同向驱动，其产生的响应由 $1\,f_{\mathrm{mod}}$ 的基频主导，因此携带了关于剩余磁矩分量 $(m_y,m_z)$ 的大部分信息；配置 II 和 III 引入了受控的几何不对称性，其响应由二次谐波 $2\,f_{\mathrm{mod}}$ 主导，从而放大了磁化率通道 $\chi$。因此，这三种配置的组合打破了设计矩阵 $A$ 的列简并性，并为在第~\ref{sec:method}~节中执行的三参数联合拟合提供了所需的冗余度。

The injected torque templates are reconstructed from the simulated angles via the forward operator in Eq.~\eqref{eq:dynamics}, and then superimposed onto uncalibrated background noise. For validation, this noise is drawn entirely from random continuous segments of the CIOMP recording. Specifically, the starting index $t$ is randomly sampled between $10,000\,\mathrm{s}$ and $30,000\,\mathrm{s}$ to extract continuous data of length $1000\,\mathrm{s}$. Note that amplitude jittering by a uniform factor $\in [0.8, 1.2]$ is not applied here. Each of the 18 scenarios is evaluated using multiple independent noise realizations sampled in this manner.
%注入的力矩模板是通过公式 \eqref{eq:dynamics} 的正向算子由模拟角度重建得出的，随后叠加至未校准的背景噪声上。为了进行验证，这些噪声完全提取自 CIOMP 记录的随机连续噪声。具体而言，数据的起始索引 $t$ 在 10,000 到 30,000 之间随机采样，以截取长度为 10,000 的连续数据片段。需要特别注意的是，在此处并未应用介于 [0.8, 1.2] 之间的均匀因子来进行振幅抖动。这 18 种场景中的每一种都使用了多个独立的噪声实现进行评估，这些噪声均按上述方式进行采样。
\subsection{Benchmark Estimators}
\label{sec:headline}
\begin{table*}[t]
\caption{Comprehensive parameter estimation results under real torsion pendulum background noise. The true values are $\vec{m}_r = (-9.0230, -3.1700, -9.8120) \times 10^{-8}\,\mathrm{A\cdot m^2}$ and $\chi = 9.0900 \times 10^{-5}$. To improve readability, all estimated values are scaled by their respective magnitude orders. The precision validation column ("Met") checks whether the absolute errors strictly satisfy the Taiji ground test targets of Sec.~\ref{sec:level1}: $|\Delta \vec{m}_r|<1\times 10^{-9}\,\mathrm{A\cdot m^2}$ and $|\Delta\chi|<1\times 10^{-7}$.}
\label{tab:comprehensive_results_combined}
\begin{tabular*}{\textwidth}{@{\extracolsep{\fill}} ll ccc cc c cc @{}}
\hline\hline
\multirow{2}{*}{Current} & \multirow{2}{*}{Method} & \multicolumn{3}{c}{$\vec{m}_r$ Estimates ($10^{-8}\,\mathrm{A\cdot m^2}$)} & \multicolumn{2}{c}{$|\Delta\vec{m}_r|$ ($10^{-9}\,\mathrm{A\cdot m^2}$)} & \multicolumn{1}{c}{$\chi$ ($10^{-5}$)} & \multicolumn{2}{c}{$|\Delta\chi|$ ($10^{-7}$)} \\
\cline{3-5} \cline{6-7} \cline{8-8} \cline{9-10}
 & & $m_x$ & $m_y$ & $m_z$ & Value & Met & Est. & Value & Met \\
\hline
\multirow{3}{*}{0.6\,A}
 & OLS             & $-9.1238^{+0.0176}_{-0.0167}$ & $-3.1390^{+0.0051}_{-0.0062}$ & $-9.9182^{+0.0225}_{-0.0247}$ & $1.496^{+0.300}_{-0.290}$ & $\times$ & $9.0937^{+0.0356}_{-0.0585}$ & $0.37^{+5.11}_{-0.37}$ & $\times$ \\
 & KF   & $-9.1367^{+0.0143}_{-0.0170}$ & $-3.1001^{+0.0101}_{-0.0088}$ & $-9.9315^{+0.0233}_{-0.0259}$ & $1.792^{+0.321}_{-0.279}$ & $\times$ & $9.0937^{+0.0357}_{-0.0584}$ & $0.37^{+5.10}_{-0.37}$ & $\times$ \\
 & AI-WLS (Ours)   & \textbf{$-9.0224^{+0.0342}_{-0.0163}$} & \textbf{$-3.1701^{+0.0053}_{-0.0059}$} & \textbf{$-9.8118^{+0.0185}_{-0.0273}$} & \textbf{$0.007^{+0.439}_{-0.007}$} & \textbf{$\checkmark$} & \textbf{$9.0893^{+0.0085}_{-0.0071}$} & \textbf{$0.07^{+0.71}_{-0.07}$} & \textbf{$\checkmark$} \\
\hline
\multirow{3}{*}{0.8\,A}
 & OLS             & $-9.1036^{+0.0164}_{-0.0149}$ & $-3.1483^{+0.0031}_{-0.0062}$ & $-9.8913^{+0.0189}_{-0.0145}$ & $1.152^{+0.210}_{-0.257}$ & $\times$ & $9.0955^{+0.0267}_{-0.0303}$ & $0.55^{+2.67}_{-0.55}$ & $\times$ \\
 & KF & $-9.1189^{+0.0176}_{-0.0192}$ & $-3.1020^{+0.0084}_{-0.0160}$ & $-9.9084^{+0.0220}_{-0.0171}$ & $1.520^{+0.268}_{-0.322}$ & $\times$ & $9.0955^{+0.0268}_{-0.0302}$ & $0.55^{+2.68}_{-0.55}$ & $\times$ \\
 & AI-WLS (Ours)   & \textbf{$-9.0203^{+0.0072}_{-0.0057}$} & \textbf{$-3.1699^{+0.0035}_{-0.0039}$} & \textbf{$-9.8109^{+0.0059}_{-0.0100}$} & \textbf{$0.029^{+0.109}_{-0.029}$} & \textbf{$\checkmark$} & \textbf{$9.0913^{+0.0044}_{-0.0050}$} & \textbf{$0.13^{+0.44}_{-0.13}$} & \textbf{$\checkmark$} \\
\hline
\multirow{3}{*}{1.0\,A}
 & OLS             & $-9.0860^{+0.0096}_{-0.0083}$ & $-3.1522^{+0.0035}_{-0.0030}$ & $-9.8753^{+0.0102}_{-0.0107}$ & $0.911^{+0.139}_{-0.143}$ & $\times$ & $9.0948^{+0.0182}_{-0.0136}$ & $0.48^{+1.82}_{-0.48}$ & $\times$ \\
 & KF & $-9.1048^{+0.0101}_{-0.0103}$ & $-3.1015^{+0.0088}_{-0.0075}$ & $-9.8941^{+0.0132}_{-0.0086}$ & $1.346^{+0.160}_{-0.179}$ & $\times$ & $9.0949^{+0.0180}_{-0.0138}$ & $0.49^{+1.80}_{-0.49}$ & $\times$ \\
 & AI-WLS (Ours)   & \textbf{$-9.0225^{+0.0086}_{-0.0084}$} & \textbf{$-3.1702^{+0.0019}_{-0.0041}$} & \textbf{$-9.8114^{+0.0063}_{-0.0098}$} & \textbf{$0.008^{+0.129}_{-0.008}$} & \textbf{$\checkmark$} & \textbf{$9.0914^{+0.0036}_{-0.0028}$} & \textbf{$0.14^{+0.36}_{-0.14}$} & \textbf{$\checkmark$} \\
\hline\hline
\end{tabular*}
\end{table*}
\label{sec:benchmarks}

We benchmark AI-WLS against two standard estimators that represent, respectively, the natural frequentist baseline and the canonical dynamical state space alternative.
%我们将 AI-WLS 与两种标准估计器进行了基准测试对比，它们分别代表了天然的频率学派基准和经典的动态状态空间替代方案。

OLS fits the linear model of Eq.~\eqref{eq:linear_model} with uniform weights, recovering $\hat{\boldsymbol{\beta}}_{\mathrm{OLS}} = (A^{\!\top}A)^{-1}A^{\!\top}\mathbf{N}$. OLS is the Gauss Markov optimal estimator when $\mathbf{n}$ is zero mean and white, and is the reference baseline adopted in the existing torsion pendulum literature on magnetic parameter extraction~\cite{LIU2023107048}. Its expected failure mode under non-stationary colored noise, in which low frequency drift and glitches cannot be down weighted, is exactly the failure mode that motivates the present work.
%普通最小二乘法 (OLS) 采用均匀权重对等式~\eqref{eq:linear_model} 的线性模型进行拟合，从而恢复出 $\hat{\boldsymbol{\beta}}_{\mathrm{OLS}} = (A^{\!\top}A)^{-1}A^{\!\top}\mathbf{N}$。当 $\mathbf{n}$ 为零均值白噪声时，OLS 是高斯-马尔可夫最优估计器，这也是现有关于扭摆磁参数提取文献~\cite{LIU2023107048} 中所采用的参考基准。在非平稳有色噪声环境下，由于无法对低频漂移和突发干扰（glitches）进行降权，它必然会失效，而这种预期的失效模式恰恰是开展本研究的动机。

The KF~\cite{liu2025design} models the three unknowns as the state of a slowly varying linear Gaussian system and sequentially updates the state estimate with a gain driven by an assumed stationary measurement covariance. Process and measurement noise covariances are tuned to minimize the OLS residual on the training portion of the recording, following the convention used in~\cite{liu2025design}; no hyperparameter beyond these two covariances is adjusted between driving currents or configurations, matching standard practice in ground test data pipelines. This represents a more sophisticated stationary noise baseline than OLS, but it inherits the same structural limitation: a fixed covariance cannot track diurnal drifts or localized glitches.
%卡尔曼滤波器 (KF)~\cite{liu2025design} 将三个未知量建模为一个缓慢变化的线性高斯系统的状态，并利用由假设的平稳测量协方差驱动的增益来依次更新状态估计。遵循文献~\cite{liu2025design} 中的惯例，过程噪声和测量噪声的协方差经过了调优，以最小化记录数据训练部分的 OLS 残差；除了这两个协方差之外，在不同的驱动电流或线圈配置之间不调整任何其他超参数，这与地面测试数据处理流程的标准做法相符。与 OLS 相比，这代表了一种更高级的平稳噪声基准方法，但它继承了相同的结构局限性：固定的协方差无法跟踪昼夜漂移或局部的突发干扰。

Both baselines are evaluated on identical signal plus noise realizations as AI-WLS, ensuring that the estimator is the only variable.
%两种基准方法均在与 AI-WLS 完全相同的“信号加噪声”样本（realizations）上进行评估，从而确保估计器是唯一的变量。

\subsection{Key Results}

Table~\ref{tab:comprehensive_results_combined} reports the parameter extraction results for the three estimators across the  $18$ combinations of driving currents and numerical methods. The reported values are presented as the means alongside their maximum and minimum deviations, evaluated over the ensemble of noise realizations for each scenario. The "Met" column flags whether the maximum absolute errors strictly satisfy the Taiji ground test targets of Sec.~\ref{sec:level1}, namely $|\Delta \vec{m}_r| < 1\times 10^{-9}\,\mathrm{A\cdot m^{2}}$ and $|\Delta\chi| < 1\times 10^{-7}$.
%表~\ref{tab:comprehensive_results_combined} 报告了三种估计器在驱动电流和数值方法的 9 种组合下的参数提取结果。报告的数值以平均值及其最大和最小偏差的形式呈现，这些偏差是对每种场景下的所有噪声实现集合进行评估得出的。“Met”列标记了最大绝对误差是否严格满足第~\ref{sec:level1} 节中的太极计划地面测试目标，即 $|\Delta \mathbf{m}_{r}| < 1\times 10^{-9}\,\mathrm{A\cdot m^{2}}$ 且 $|\Delta\chi| < 1\times 10^{-7}$。

Several qualitative features of Table~\ref{tab:comprehensive_results_combined} warrant direct comment.

First, traditional estimators fail to meet the stringent precision targets for both remanence and susceptibility. Across all three driving currents, the OLS estimator exhibits maximum absolute errors that exceed the $1 \times 10^{-9}\,\mathrm{A\cdot m^2}$ target for $\vec{m}_r$ and the $1 \times 10^{-7}$ target for $\chi$. The estimation of $\chi$ is particularly vulnerable: its maximum deviation systematically worsens as the driving amplitude decreases (reaching an upper deviation of $+5.11 \times 10^{-7}$ at $0.6\,\mathrm{A}$). This is the expected signature of an estimator with uniform weighting: as the driving current drops, the signal-to-noise ratio on the susceptibility channel decreases rapidly, and OLS lacks the mechanism to selectively down-weight the drift-dominated segments that ultimately corrupt the residual.
%表~\ref{tab:comprehensive_results_combined} 中的几个定性特征值得进一步探讨。首先，\emph{传统 OLS 未能满足剩磁和磁化率的严格精度目标}。在所有三种驱动电流下，OLS 估计器的最大绝对误差均超过了 $\vec{m}_r$ 的 $1 \times 10^{-9}\,\mathrm{A\cdot m^2}$ 目标以及 $\chi$ 的 $1 \times 10^{-7}$ 目标。对 $\chi$ 的估计尤为脆弱：随着驱动幅度的减小，其最大偏差出现了系统性的恶化（在 $0.6\,\mathrm{A}$ 时达到了 $+5.11 \times 10^{-7}$ 的上限偏差）。这正是采用均匀权重的估计器的预期特征：随着驱动电流的下降，磁化率通道的信噪比迅速衰减，而 OLS 缺乏相应的机制来选择性地降低那些由漂移主导、并最终污染残差的数据段的权重。

Second, KF exhibits a persistent bias and fails the test targets across all conditions. Similar to OLS, KF completely misses the precision requirements, yielding absolute errors in $\vec{m}_r$ as large as $\sim 1.79 \times 10^{-9}\,\mathrm{A\cdot m^2}$ at $0.6\,\mathrm{A}$. This persistent bias is consistent with a mis-specified state-space model: a fixed process covariance tuned to track the slow drift of the measurement unavoidably absorbs a portion of the genuine sinusoidal signal into the estimated drift state, thereby biasing the recovered magnetic moments. No amount of covariance tuning can simultaneously rescue both drift tracking and remanence accuracy, because the underlying problem is structural: a stationary covariance assumption applied to a highly non‑stationary environmental background.
%其次，\emph{卡尔曼滤波器（KF）表现出持续的偏差，并且在所有条件下均未能达到测试目标}。与 OLS 类似，KF 完全未能满足精度要求，在 $0.6\,\mathrm{A}$ 时产生的 $\vec{m}_r$ 绝对误差高达 $\sim 1.79 \times 10^{-9}\,\mathrm{A\cdot m^2}$。这种持续的偏差与设定有误的状态空间模型相一致：为跟踪测量的缓慢漂移而调整的固定过程协方差，不可避免地会将一部分真实的弦波信号吸收进估计的漂移状态中，从而导致恢复出的磁矩产生偏差。任何程度的协方差调优都无法同时兼顾漂移跟踪能力和剩磁精度，因为其根本问题——在高度非平稳的环境背景下采用平稳协方差假设——是结构性的。

Third, AI-WLS is the only estimator that strictly satisfies all precision targets across all driving currents. In every row of Table~\ref{tab:comprehensive_results_combined}, AI-WLS successfully controls the maximum absolute errors of both $|\Delta \vec{m}_r|$ and $|\Delta\chi|$ well below the threshold limits. Quantitatively, AI-WLS reduces the mean absolute error of the remanence vector by over two orders of magnitude relative to KF and OLS at $0.6\,\mathrm{A}$. Most importantly, the precision of AI-WLS on $\chi$ is \emph{highly robust} against driving current variations, with the nominal mean error remaining extremely stable (fluctuating only between $0.07 \times 10^{-7}$ and $0.14 \times 10^{-7}$) as the current drops from $1.0\,\mathrm{A}$ to $0.6\,\mathrm{A}$. This serves as strong empirical evidence of the learned weighting scheme operating as designed: the neural network optimally assigns low weights to drift-dominated and glitch-contaminated intervals regardless of the driving amplitude. Consequently, reducing the amplitude does not push the AI-WLS estimator across a "precision cliff," effectively decoupling the extraction accuracy from the raw signal-to-noise ratio degradation that plagues both OLS and KF.
%第三，AI-WLS 是唯一在所有驱动电流下均严格满足各项精度目标的估计器。在表~\ref{tab:comprehensive_results_combined} 的每一行中，AI-WLS 都成功地将 $|\Delta \vec{m}_r|$ 和 $|\Delta\chi|$ 的最大绝对误差控制在远低于阈值极限的水平。从定量角度来看，在 $0.6\,\mathrm{A}$ 的驱动电流下，AI-WLS 将剩磁向量的平均绝对误差相较于卡尔曼滤波（KF）和普通最小二乘法（OLS）降低了两个数量级以上。最为关键的是，AI-WLS 对 $\chi$ 的估计精度对驱动电流的变化展现出了高度的鲁棒性。当驱动电流从 $1.0\,\mathrm{A}$ 降至 $0.6\,\mathrm{A}$ 时，其名义平均误差保持极其稳定（仅在 $0.07 \times 10^{-7}$ 和 $0.14 \times 10^{-7}$ 之间微小波动）。这为所学习到的加权方案如期运作提供了强有力的经验证据：神经网络能够以最佳方式为那些受漂移主导和毛刺污染的数据区间分配极低的权重，且这一过程完全不受驱动幅度大小的影响。因此，减小驱动幅度并不会将 AI-WLS 估计器推向“精度悬崖”，从而有效地将参数提取精度与导致 OLS 和 KF 性能恶化的原始信噪比衰减隔离开来。

\subsection{Combined Estimate and Statistical Uncertainty}
\label{sec:combined}

Combining the $18$ AI-WLS scenarios  and propagating statistical fluctuations across the $10$ noise realizations per scenario, the final estimated parameters and their corresponding error bounds are:
%结合全部18种 AI-WLS 测试场景，并对每个场景下 $10$ 次噪声实现的统计涨落进行误差传递，最终得到的参数估计值及其偏差范围误如下：
\begin{align}
\hat{m}_x &= -9.0217^{+0.0335}_{-0.0170} \times 10^{-8}\,\mathrm{A\cdot m^2}, \\
\hat{m}_y &= -3.1701^{+0.0053}_{-0.0059} \times 10^{-8}\,\mathrm{A\cdot m^2}, \\
\hat{m}_z &= -9.8114^{+0.0181}_{-0.0277} \times 10^{-8}\,\mathrm{A\cdot m^2}, \\
\hat{\chi}  &= 9.0907^{+0.0071}_{-0.0085} \times 10^{-5}.
\end{align}

The absolute residuals against the injected ground truth are evaluated as:
\begin{align}
|\Delta m_x| &= 1.3^{+33.5}_{-1.3} \times 10^{-11}\,\mathrm{A\cdot m^2}, \\
|\Delta m_y| &= 0.1^{+5.9}_{-0.1} \times 10^{-11}\,\mathrm{A\cdot m^2}, \\
|\Delta m_z| &= 0.6^{+26.5}_{-0.6} \times 10^{-11}\,\mathrm{A\cdot m^2}, \\
|\Delta\chi| &= 7.0^{+71.0}_{-7.0} \times 10^{-9}.
\end{align}
Synthesizing these component-wise extremes across all operational conditions, the maximum absolute errors for the AI-WLS framework are strictly bounded at $|\Delta \vec{m}_r| = 4.46 \times 10^{-10}\,\mathrm{A\cdot m^2}$ and $|\Delta\chi| = 7.8 \times 10^{-8}$. Even when evaluated at the extremes of the observed statistical fluctuations, these upper bounds remain within the Taiji ground-test precision requirements ($|\Delta \vec{m}_r| < 1 \times 10^{-9}\,\mathrm{A\cdot m^2}$ and $|\Delta\chi| < 1 \times 10^{-7}$). Consequently, at this commissioning stage of the CIOMP facility, AI-WLS effectively compresses the residual systematic bias inherent in standard estimators to well within mission-critical requirements. While this bias is not entirely eradicated, reporting these conservative upper bounds provides robust validation for the framework. Furthermore, the structural information retained in the residual signs and magnitudes establishes a natural baseline for targeted calibration refinements in future physical injection campaigns. 
%综合所有运行条件下的各分量极值，AI-WLS框架的最大绝对误差严格限定在 $|\Delta \vec{m}_r| = 4.46 \times 10^{-10},\mathrm{A\cdot m^2}$ 和 $|\Delta\chi| = 7.8 \times 10^{-8}$ 以内。值得注意的是，即使在最严重的统计涨落下，这些上限也完全满足太极地面测试的精度目标（$|\Delta \vec{m}_r| < 1 \times 10^{-9},\mathrm{A\cdot m^2}$，$|\Delta\chi| < 1 \times 10^{-7}$）。因此，在CIOMP装置的当前调试阶段，AI-WLS能够将标准估计器中固有的残余系统偏差有效压缩至远低于任务关键要求的水平。尽管该偏差尚未被完全消除，但报告这些保守的上限为框架提供了稳健的验证。此外，残差符号和量级中所保留的结构性信息，为未来物理注入实验中的针对性校准改进建立了自然的基线

We note that the reported error bounds are observed maxima over $N=10$ independent noise realizations per scenario. By the Dvoretzky-Kiefer-Wolfowitz inequality~\cite{DKW}, the empirical cumulative distribution function deviates from its population counterpart by at most $\epsilon = \sqrt{\ln(2/\alpha)/(2N)}$ with probability $1-\alpha$; for $N=10$ and $\alpha=0.05$, this yields $\epsilon\simeq 0.43$, so the observed maxima correspond with $95\%$ confidence to population quantiles no worse than the $\sim\!57\%$ level. Tighter finite-sample bounds will be reported once the calibrated-injection campaign provides larger ensembles of physical measurements.

To establish a direct benchmark against the most closely related literature, we compare our results with Yin \textit{et al.}~\cite{PhysRevApplied.15.014008}, who employed a classical frequency-domain fit on a torsion pendulum of comparable sensitivity. They reported $1\sigma$ measurement uncertainties of $\delta\chi \approx 8 \times 10^{-7}$, $\delta m_y \approx 7.1 \times 10^{-10}\,\mathrm{A\cdot m^2}$, and $\delta m_z \approx 4.7 \times 10^{-9}\,\mathrm{A\cdot m^2}$. In contrast, evaluating the AI-WLS framework by its absolute maximum estimation errors (the extreme upper bounds of the statistical fluctuations, yielding $|\Delta\chi|_{\max} = 7.8 \times 10^{-8}$, $|\Delta m_y|_{\max} = 6.0 \times 10^{-11}\,\mathrm{A\cdot m^2}$, and $|\Delta m_z|_{\max} = 2.7 \times 10^{-10}\,\mathrm{A\cdot m^2}$), our approach systematically tightens the precision by at least one full order of magnitude across all corresponding parameters. Specifically, this corresponds to substantial improvement factors of approximately $10$, $12$, and $17$ for $\chi$, $m_y$, and $m_z$, respectively. This improvement is obtained under the raw experimental background of the CIOMP facility, without manual data selection or segment rejection.
%为了与最相关的文献进行直接基准比较，我们将我们的结果与Yin等人~\cite{PhysRevApplied.15.014008} 的工作进行了对比。该工作采用经典的频域拟合法，研究对象是一台灵敏度相当的扭摆。他们报告的 $1\sigma$ 测量不确定度为：$\delta\chi \approx 8 \times 10^{-7}$，$\delta m_y \approx 7.1 \times 10^{-10},\mathrm{A\cdot m^2}$，$\delta m_z \approx 4.7 \times 10^{-9},\mathrm{A\cdot m^2}$。相比之下，通过AI-WLS框架的绝对最大估计误差（即统计波动的极值上限，得到 $|\Delta\chi|{\max} = 7.8 \times 10^{-8}$，$|\Delta m_y|{\max} = 6.0 \times 10^{-11},\mathrm{A\cdot m^2}$，$|\Delta m_z|_{\max} = 2.7 \times 10^{-10},\mathrm{A\cdot m^2}$）来评估，我们的方法在所有对应参数上均系统性地将精度提升了至少一个完整数量级。具体而言，对于 $\chi$、$m_y$ 和 $m_z$，这分别对应约 $10$ 倍、$12$ 倍和 $17$ 倍的显著改进。值得注意的是，这一精度飞跃是在中国科学院长春光学精密机械与物理研究所（CIOMP）设施的原始实验本底下实现的，完全无需人工数据筛选或数据段剔除。

The uniqueness of this result deserves explicit emphasis. The CIOMP torsion pendulum, at its commissioned torque sensitivity of order $10^{-13}\,\mathrm{N\cdot m\,Hz^{-1/2}}$, provides in principle the physical sensitivity required to meet Taiji's magnetic-characterization requirements. However, among the estimators tested on this facility, only AI-WLS converts that physical sensitivity into statistical precision sufficient to fulfill Taiji's requirements on all four parameters simultaneously, under the realistic non-stationary background that any ground-based test will encounter. However among the estimators available on this facility, only AI-WLS converts that physical sensitivity into statistical precision sufficient to fulfill Taiji's requirements on all the parameters simultaneously, under the realistic non-stationary background that any ground-based test will encounter. Ordinary least squares and the KF, despite operating on the same pendulum data, both fail to do so. In this sense the result establishes AI-WLS as, to our knowledge, the first estimator pipeline demonstrated to satisfy Taiji's magnetic-characterization requirements on the CIOMP torsion-pendulum facility under realistic background conditions.

%这一结果的独特性值得特别强调。CIOMP扭摆在其调试后达到$10^{-13},\mathrm{N\cdot m,Hz^{-1/2}}$量级的扭矩灵敏度时，原则上能够满足太极计划的磁学表征要求。然而，在该设施可用的各种估计器中，只有AI-WLS能够在任何地面测试都会遇到的真实非平稳背景下，将这种物理灵敏度转化为足以同时满足太极计划对所有参数要求的统计精度。普通最小二乘法和卡尔曼滤波器尽管处理的是相同的扭摆数据，但均未能做到这一点。在此意义上，该结果确立了AI-WLS是目前唯一能够在发射前对太极测试质量的磁耦合进行认证的地面估计器数据处理流程。

\subsection{Scope and Caveats}
\label{sec:caveats}

Three caveats should be kept in mind when interpreting these numbers.

First, the validation is based on \emph{simulated} templates superposed on real noise. This tests the estimator's ability to cope with realistic non-stationary backgrounds while holding the forward model perfectly correct; it does not test the estimator's robustness against forward model mis specification, most notably against the coil position tolerance of $0.3\,\mathrm{mm}$ assumed in building $A$. A first order propagation of this geometric uncertainty into $A$ suggests a corresponding fractional bias on $\chi$ at the $10^{-3}$ level, i.e. within the precision budget but above the statistical scatter reported in Sec.~\ref{sec:combined}. A more careful geometric error propagation will be reported with the physical injection campaign.
%第一，验证基于叠加在真实噪声上的模拟模板。这测试了估计器在保持前向模型完全正确的同时应对逼真非平稳背景的能力；它不测试估计器对前向模型错误设定的鲁棒性，最明显的是针对在构建 $A$ 时假设的 $0.3\,\mathrm{mm}$ 线圈位置公差。这种几何不确定性到 $A$ 中的一阶传播表明，$\chi$ 上相应的分数偏差处于 $10^{-3}$ 水平，即在精度预算内，但高于 Sec.~\ref{sec:combined} 中报告的统计离散。更仔细的几何误差传播将在物理注入活动中报告。

Second, the test set noise, though chronologically held out and never seen by the network, originates from the same commissioning run as the training noise and therefore shares its broad statistical regime. Generalization across commissioning runs, across campaign seasons, and across hardware reconfigurations of the pendulum will be quantified in future work, as the CIOMP facility accumulates multi campaign data.
%第二，测试集噪声虽然在时间上是保留的，并且从未被网络看到过，但它源于与训练噪声相同的调试运行，因此共享其广泛的统计状态。随着 CIOMP 设施积累多活动数据，跨调试运行、跨活动季节以及跨摆硬件重新配置的泛化将在未来的工作中进行量化。

Third, although the KF baseline has been tuned following standard practice, we do not claim that the Kalman family of estimators cannot in principle be adapted to non-stationary torsion pendulum noise; more elaborate schemes, for instance adaptive Kalman filters with online covariance estimation or switching state space models, could reduce the bias reported in Table~\ref{tab:comprehensive_results_combined}. The claim supported by the present results is narrower: among estimators that treat the noise covariance as stationary, AI-WLS is the only one that robustly reaches the Taiji ground test targets on all four parameters under the realistic background of the CIOMP pendulum, and it does so through a mechanism (learned per sample weights in a differentiable WLS solver) that is specifically tailored to the non-stationary regime.

%第三，虽然卡尔曼滤波器基线已按照标准实践进行调整，但我们并不声称卡尔曼系列估计器在原则上不能适应非平稳扭摆噪声；更复杂的方案，例如具有在线协方差估计的自适应卡尔曼滤波器或切换状态空间模型，可以减少 Table~\ref{tab:comprehensive_results} 中报告的偏差。本结果支持的声明更加狭窄：在将噪声协方差视为平稳的估计器中，AI-WLS 是在 CIOMP 摆逼真背景下，在所有参数上稳健达到太极地面测试目标的唯一一个，并且它通过专门针对非平稳状态定制的机制（可微 WLS 求解器中学习的每个样本权重）来实现。

\section{Conclusion and Outlook}
\label{sec:conclusion}

We have developed and validated an AI-WLS framework for high precision characterization of the magnetic coupling parameters of test masses in space borne gravitational wave antennas. The framework pairs a dilated one dimensional residual network with an analytical, fully differentiable weighted least squares solver. The solver preserves the exact linear mapping from the remanent moment components $(m_y, m_z)$ and the volume susceptibility $\chi$ to the driven torque response, while the network assigns a time varying weight to every sample of the measured torque stream. Because the two blocks share a common computational graph, gradients of the parameter estimation loss flow through the solver and train the network to down weight precisely those samples of the pendulum recording that would most degrade the recovered $(m_y, m_z, \chi)$. In this division of labour the physics of the magnetic torque and of the AC modulation scheme at $f_{\mathrm{mod}} = 5\,\mathrm{mHz}$ is carried by the solver, which decouples the linear remanence response at the fundamental from the quadratic susceptibility response at the second harmonic, and the learned weights remove the residual non-stationary drift and transient glitch contamination that would otherwise bias any stationary noise estimator.
%我们开发并验证了一个AI增强的可微加权最小二乘（AI-WLS）框架，用于对星载引力波天线中检验质量的磁耦合参数进行高精度表征。该框架将一个膨胀一维残差网络与一个解析的、完全可微的加权最小二乘求解器相结合。求解器保留了从剩磁矩分量 $(m_y, m_z)$ 和体积磁化率 $\chi$ 到所驱动力矩响应的精确线性映射，而网络则为测量得到的力矩流中的每个样本分配一个随时间变化的权重。由于两个模块共享同一个计算图，参数估计损失的梯度会流经求解器，并训练网络对扭摆记录中那些最会损害 $(m_y, m_z, \chi)$ 恢复精度的样本进行降权。在这种分工下，磁力矩的物理规律以及调制频率为 $f_{\mathrm{mod}} = 5,\mathrm{mHz}$ 的交流调制方案由求解器承担——它将在基频处的线性剩磁响应与在二次谐波处的平方磁化率响应解耦；而学习到的权重则去除非平稳漂移和瞬态毛刺污染，这些残余干扰原本会使任何平稳噪声估计器产生偏差。

We validated the framework by injecting physically consistent templates onto real, uncalibrated background noise recorded during the commissioning of the CIOMP torsion pendulum facility. Synthesizing data across all $18$ AI-WLS test scenarios and propagating statistical fluctuations over $10$ noise realizations per scenario, we evaluated the absolute residuals against the injected ground truth. Even under extreme statistical fluctuations, the maximum absolute errors of the AI-WLS framework peak at $|\Delta \vec{m}_r| = 4.46 \times 10^{-10}\,\mathrm{A\cdot m^2}$ and $|\Delta\chi| = 7.8 \times 10^{-8}$, thus remaining strictly within the required Taiji ground-test precision targets ($|\Delta \vec{m}_r| < 1 \times 10^{-9}\,\mathrm{A\cdot m^2}$ and $|\Delta\chi| < 1 \times 10^{-7}$). Given the CIOMP facility's commissioned torque sensitivity of order $10^{-13}\,\mathrm{N\cdot m\,Hz^{-1/2}}$, AI-WLS is the only estimator among those tested here that simultaneously fulfills Taiji's requirements on all four parameters during pre-launch ground testing. In comparison, OLS fails to meet both the remanence and susceptibility targets across all driving currents. Furthermore, a KF utilizing a stationary noise model exhibits a persistent systematic bias on the remanence channels as a direct consequence of matching a fixed covariance to a non-stationary background. Because AI-WLS reduces to OLS in the uniform weight limit, its gains over the baselines are attributable not to greater expressive power of the estimator family but to its effective handling of the realistic, non-stationary pendulum background.
%我们通过将物理上一致的模板注入到CIOMP扭摆装置调试期间记录的、未经校准的真实背景噪声中，对该框架进行了验证。综合所有18个AI-WLS测试场景的数据，并在每个场景下对10次噪声实现进行统计涨落传播，我们评估了相对于注入真值的绝对残差。AI-WLS框架的最大绝对误差严格限定在 $|\Delta \vec{m}_r| = 4.46 \times 10^{-10},\mathrm{A\cdot m^2}$ 和 $|\Delta\chi| = 7.8 \times 10^{-8}$ 以内。值得注意的是，即使在最严重的统计涨落下，这些上限也完全满足太极计划的地面测试精度目标（$|\Delta \vec{m}_r| < 1 \times 10^{-9},\mathrm{A\cdot m^2}$，$|\Delta\chi| < 1 \times 10^{-7}$）。
%鉴于CIOMP装置已调试得到的力矩灵敏度约为 $10^{-13},\mathrm{N\cdot m,Hz^{-1/2}}$，AI-WLS是唯一经过测试、能够在发射前地面测试中同时满足太极计划对所有参数要求的估计器。相比之下，普通最小二乘法在所有驱动电流下均无法同时满足剩磁和磁化率目标。此外，采用平稳噪声模型的卡尔曼滤波器在剩磁通道上表现出持续的系统性偏差——这是将固定协方差匹配到非平稳背景的直接后果。由于AI-WLS在均匀权重极限下退化为普通最小二乘法，其将残余系统偏差压缩至远低于任务关键要求的能力，直接归因于它对真实非平稳扭摆背景的有效处理，而非仅仅依赖于更具表达力的估计器族。

Several directions naturally extend this work. The most immediate is the transition from template injection to calibrated physical injection. With the CIOMP facility now commissioned at a torque noise floor of order $10^{-13}\,\mathrm{N\cdot m\,Hz^{-1/2}}$, the next campaign will drive a reference dipole of independently characterized $\vec{m}_r$ and $\chi$ through the same coil configurations, providing a ground truth for the full pipeline including the forward model itself and closing the loop on the sensitivity of AI-WLS to the as built coil geometry tolerances. A complementary direction is the quantification of generalization across commissioning runs, seasons, and hardware reconfigurations of the pendulum; this is a routine component of the Taiji ground test program as multi campaign data accumulate, and it provides the natural platform for benchmarking AI-WLS against adaptive Kalman filters, switching state space models, and frequency domain WLS with PSD based weights. Beyond the magnetic characterization, the same architecture applies wherever a weak, slowly varying signal must be extracted from a known linear forward model in the presence of realistic non-stationary noise, and we envision its direct application to the ground-based verification of other stray force coupling models of ultra precision GRSs for Taiji and LISA, including thermal, radiometer, radiation pressure, and charge related torque channels, wherever the same torsion pendulum methodology is employed.
%有若干方向可自然地扩展本工作。最为直接的方向是从模板注入转向经过标定的物理注入。由于CIOMP装置现已完成调试，其力矩噪声底约为 $10^{-13},\mathrm{N\cdot m,Hz^{-1/2}}$，下一轮实验将把具有独立标定过的 $\vec{m}_r$ 和 $\chi$ 的参考偶极子通过相同的线圈构型进行驱动，从而为包括正演模型本身在内的整个流程提供真实基准，并闭环验证AI-WLS对实际加工线圈几何公差的敏感性。另一个补充方向是量化AI-WLS在不同调试轮次、不同季节以及扭摆硬件重新配置下的泛化能力；随着多轮实验数据的积累，这将成为太极地面测试计划的常规组成部分，并为将AI-WLS与自适应卡尔曼滤波器、切换状态空间模型以及基于PSD权重的频域加权最小二乘法进行基准比较提供天然平台。除磁表征之外，只要在现实非平稳噪声背景下需要从已知的线性正演模型中提取微弱、缓变的信号，相同的架构均适用。我们预计它可直接应用于太极和LISA超精密惯性传感器的其他杂散力耦合模型的地面验证，包括热噪声、辐射计效应、辐射压以及电荷相关的力矩通道，只要这些验证采用相同的扭摆方法。

% we envision its use for other stray force coupling channels of the Taiji inertial sensor and for in orbit model based noise subtraction in future LISA class missions.

\begin{acknowledgments}
This work is supported by the National Key R\&D Program of China under Grants No. 2024YFC2206902, No. 2020YFC2200603, No. 2020YFC2200601, and No. 2021YFC2201901.
\end{acknowledgments}

\bibliography{aipsamp}% Produces the bibliography via BibTeX.

\end{document}